\documentclass[pdflatex,sn-mathphys-num]{sn-jnl}

\usepackage{graphicx}%
\usepackage{multirow}%
\usepackage{amsmath,amssymb,amsfonts}%
\usepackage{amsthm}%
\usepackage{mathrsfs}%
\usepackage[title]{appendix}%
\usepackage{xcolor}%
\usepackage{textcomp}%
\usepackage{manyfoot}%
\usepackage{booktabs}%
\usepackage{algorithm}%
\usepackage{algorithmicx}%
\usepackage{algpseudocode}%
\usepackage{listings}%
\usepackage{amssymb}
\usepackage{enumitem}

\theoremstyle{thmstyleone}%
\theoremstyle{thmstyletwo}%

\theoremstyle{thmstylethree}%

\begin{document}

\title[Perception of BCI Implantation in SCI and Stroke]{Perception of Brain-Computer Interface Implantation Surgery for Motor, Sensory, and Autonomic Restoration in Spinal Cord Injury and Stroke}


\author*[1]{\fnm{Derrick} \sur{Lin}}\email{DRLin@mednet.ucla.edu, and@uci.edu}

\author[2]{\fnm{Tracie} \sur{Tran}}
\author[2]{\fnm{Shravan} \sur{Thaploo}}
\author[2]{\fnm{Jose Gabrielle E.} \sur{Matias}}
\author[3]{\fnm{Joy E.} \sur{Pixley}}
\author[2]{\fnm{Zoran} \sur{Nenadic}}
\author*[2,4]{\fnm{An H.} \sur{Do}}

\affil*[1]{\orgname{David Geffen School of Medicine at UCLA}, \orgaddress{\city{Los Angeles}, \state{CA}, \country{USA}}}

\affil[2]{\orgdiv{Department of Biomedical Engineering}, \orgname{University of California, Irvine}, \orgaddress{\city{Irvine}, \state{CA}, \country{USA}}}

\affil[3]{\orgname{Independent Researcher (formerly at University of California, Irvine)}, \orgaddress{\country{USA}}}

\affil[4]{\orgdiv{Department of Neurology}, \orgname{University of California, Irvine}, \orgaddress{\city{Irvine}, \state{CA}, \country{USA}}}


\abstract{Stroke and spinal cord injury (SCI) are neurological injuries that can significantly impact the quality of life of survivors in both the physical and psychosocial domains. Both diseases often result in significant motor and sensory impairments that are not fully reversible despite current available therapies. Invasive Brain-computer interface (BCI) technology has emerged as a promising means to bypass the site of injury and potentially restore motor and sensory function; However, to maximize the utility and participant satisfaction with such technology, participants' willingness to embrace BCI technology must be assessed, and placed in context with functional goals and rehabilitative priorities. To this end, we conducted a survey of a cohort (n=71) of stroke (n=33), SCI (n=37), and both stroke and SCI (n=1) participants regarding their receptiveness to invasive ECoG-based BCI technology as well as to assess their goals for functional rehabilitation. Overall, participants indicated a high level of willingness to undergo surgery to implant electrocorticography (ECoG) grids for BCI technology if basic motor functions, including upper extremity, gait, bowel/bladder, and sensory function were restored. There was no correlation between participant willingness to undergo a prospective BCI implantation and the level of functional recovery offered by the BCI. Similarly, there was no correlation between willingness to undergo surgery and the participants' perceived rehabilitative priorities and level of disability. These findings indicate that participants were interested in invasive BCI technology even if only basic functions can be restored, regardless of their level of disability and their rehabilitative priorities. Such observations imply that first generation commercial invasive BCIs may not need extensive functions to garner adoption. Conversely, it also raises a concern that participants from the stroke and SCI cohort may be overly enthusiastic about such technology, which poses potential risks for medical exploitation.}




\maketitle

\section{Introduction}

Stroke and spinal cord injury (SCI) are debilitating neurological conditions with no current means to reverse functional loss fully. Beyond motor impairment, patients often experience complications such as neurogenic bladder and bowel, urinary tract infections, spasticity, cardiovascular issues, and depression. Given the significant public health impact, interventions that enhance independence and address psychological effects are urgently needed. Brain-computer interface (BCI) technology has recently gained attention as a promising approach to restoring motor, sensory, and possibly autonomic functions in these populations.

BCIs are systems that translate brain signals into commands for external effectors \cite{shih2012brain}. In essence, BCIs can potentially enable participants affected by neurological injuries to gain ``brain control'' of external assistive systems such as a prosthetic limb. Currently, there are several methods for signal acquisition, including scalp electroencephalography (EEG), microelectrode arrays (MEAs), and electrocorticography (ECoG). In this study, we focused on ECoG-based BCI due to the growing utilization of ECoG in implantable BCI systems with potentially high clinical impact, such as in language/speech BCIs \cite{komeiji2024feasibility, liu2023decoding, metzger2023high}, in restoring gait after spinal cord injury (SCI) \cite{lorach2023walking, lim2024early}, and in restoring sensation \cite{lim2024early, lim2025real}. Furthermore, studies have shown that fully implantable ECoG grids exhibit long-term biocompatibility, without cortical damage and sustained neural activity recording \cite{degenhart2016histological}. The development of fully implantable and wireless BCI systems may maximize recipient functional independence while long-term biocompatibility would reduce the necessity of repeat procedures or device failure, making ECoG a promising approach for BCI signal acquisition. 

Technological potential alone does not ensure adoption. As invasive BCIs approach clinical use, understanding recipients’ perspectives is essential. Success depends not only on technical efficacy but also on users’ willingness to undergo surgery and their perceptions of potential benefits. Prior surveys suggest that most participants were aware of BCIs and held positive views, even toward invasive systems \cite{sattler2022public, monasterio2021attitudes, schmid2021thoughts}. While positive attitudes support adoption \cite{konig2022user}, user satisfaction often hinges on whether personal needs are met \cite{martin2011impact}. Although ECoG-based BCIs are technically feasible, further research is needed to assess how well they align with users’ goals for functional restoration. Ultimately, success depends on understanding end-user perceptions—not just developer enthusiasm.

Previous studies have identified key rehabilitation priorities for individuals with stroke and spinal cord injury (SCI), including motor, autonomic, and cognitive functions \cite{anderson2004targeting, brown2002consumer, simpson2012health, pollock2012top}. While these findings highlight broad functional goals, several aspects of stakeholder perception remain underexplored. Specifically, no studies to date have examined how the severity of disability or the perceived potential for functional restoration affects a user’s willingness to undergo invasive BCI surgery. Additionally, while motor outcomes are often emphasized, the potential of BCI systems to restore sensory function—and how this is perceived by users—has received limited attention.

In this study, we aimed to elucidate the receptiveness of stroke and SCI participants to invasive ECoG-based BCI technologies and assess their willingness to undergo a prospective surgery for device implantation based on potential motor and sensory restoration. Furthermore, we examined how disability severity, perceived potential functional gains, and rehabilitation priorities influence this willingness. Moreover, this survey also assessed participant attitudes toward both sensory and motor restoration. Finally, we also sought to understand concerns that stroke and SCI participants may have regarding the invasive ECoG-based BCI technology.

\section{Methods}

\subsection{Participant Recruitment}

Participants were recruited at the University of California, Irvine (UCI) neurology clinics, UCI occupational and physical therapy clinics, community rehabilitation centers, SCI or stroke outreach events, or via email and support group forums. Inclusion criteria included age $>$18 and the presence of chronic ($>$6 months post onset) SCI, stroke, or both. Exclusion criteria included having an injury for less than 6 months or not having the capability to answer the survey questions. Participants were provided with a \$10 Amazon gift card at the conclusion of the survey as an incentive. Participants completed the survey electronically through a computer. 

\subsection{Survey Development and Structure}
The survey was designed to determine 1) the importance of functional restoration to SCI and stroke participants at various levels of injury; 2) whether SCI and stroke participants at various levels of injury would consider implantation of an ECoG-based BCI device if varying degrees of motor and sensory functions can be regained; 3) whether prior knowledge of invasive BCI systems would affect the willingness to undergo surgery for a given restoration. The survey was implemented as an internet-based webform with a combination of multiple-choice and free-response questions, and was designed to take approximately 15 minutes to complete. Participants either completed the survey remotely or on-site using a provided computer, with an administrator present. To avoid introducing bias in participant responses, survey administrators did not answer any questions outside of technical issues and basic explanations of the questions and/or response options while the participant completed the survey. Participants were allowed to ask questions about BCI technology after completing the survey. Remote participants were given a weblink to respond to the survey. There were five parts to the survey: \textbf{Part 1} (question 1--6) inquired about participants' neural injury, including the type and severity of the injury; \textbf{Part 2} (question 7--8) established the participant's pre-existing familiarity with BCI technology; \textbf{Part 3} (question 9--23) determined the participant's willingness to undergo implantation of a BCI device given that certain neurological function could be restored; \textbf{Part 4} (question 24--29) inquired about additional functions that participants would like restored beyond those previously mentioned; it also assessed the likelihood of undergoing surgery to implant a BCI system for these functions and gathered any concerns regarding the implantation process; \textbf{Part 5} (question 30--35) collected demographic information, e.g., age, gender, and other social factors. Each part is described in further detail below. A copy of the survey can be found in the appendix \ref{sec:survey}. This study was reviewed by the University of California, Irvine Institutional Review Board (IRB) and determined to be exempt. Written informed consent was obtained from all participants prior to participation, in accordance with institutional and federal guidelines.

To minimize unusable data, participants were given the explicit option to select “Don’t know/Decline to answer” for most questions. Where appropriate, skip logic was used to route participants past questions irrelevant to their injury type (e.g., participants indicating paraplegia were not asked about arm function). If a participant skipped a question without selection, the response was coded as ``missing.'' Questions skipped due to programmed logic were coded as ``not applicable'' (NA). These categories (DK/REF, missing, NA) were tracked separately in the analysis.

\textbf{Part 1} of the survey was designed to establish participants' type of neural injury (SCI, stroke, or both), the extent of the injury (if known, ASIA Scale for SCI, Modified Rankin Scale for stroke), and the type of prostheses participants were currently using. This was accomplished with multiple choice questions for each of the above with the inclusion of ``Don’t know/Decline to Answer'' option to prevent inaccurate guesses if participants were uncertain. Responses marked as ``Don’t know/Decline to answer'' were retained but excluded from analyses that specifically required those data points (e.g., ASIA or Rankin stratification).

\textbf{Part 2} of the survey asked participants whether they had prior familiarity with ECoG BCI technology to determine whether prior knowledge of BCI impacted the participants' responses regarding their willingness to undergo surgery for a certain functional restoration. Regardless of the response, participants were given written and visual descriptions of a hypothetical ECoG-based BCI implant (see supplementary material for complete details) and the mechanism by which it can potentially restore function. It then detailed the necessity for the surgical implantation of an ECoG electrode grid to record brain wave activities so that all participants understand the procedure involved. 

\textbf{Part 3} of the survey was intended to establish participants' perception on the importance of regaining upper extremity motor, gait, and sensation function, as well as whether the participants were willing to undergo BCI implantation procedures if varying degrees of each of the above functions can be restored. For upper extremity function, the increments of function restored included ``basic grasp and release,'' ``fine control of fingers along with basic grasp and release'' and ``fine control of arms'' along with the previous two functions. For walking, increments included ``Ability to stand,'' ``Ability to walk at constant speed,'' ``Ability to walk at various speeds,'' and ``Ability to make turns along with walking at various speeds.'' Finally, participants were asked about their willingness to undergo surgery to restore a particular sensation, specifically regarding legs, arms, hands and fingers, and bladder fullness.

\textbf{Part 4} of the survey asked if there were any other neurological functions not mentioned in the survey that the participant would like restored and the likelihood they would undergo surgery to implant a BCI system for such functions. Part 4 also inquired about any concerns participants had with a prospective implantation of the BCI system. 

\textbf{Part 5} of the survey was composed of demographic questions (i.e., age, gender, level of education, current occupation, current living situation, and household income). 

\subsection{Analytic Methods}
\subsubsection{Participant Classification}
Responses to Part 1 (questions 1–6) of the survey—covering injury type (SCI, stroke, or both), nature of impairment (e.g., tetraplegia/paraplegia), ASIA Impairment Scale, level of injury, Modified Rankin Scale, and current mobility aid—served as the basis for classifying participants into appropriate subgroups for the analyses presented in subsequent sections.

Participants who selected “Don’t know/Decline to answer” were excluded only from analyses where their response was required for subgrouping (e.g., ASIA or Rankin classification). Participants who were programmatically skipped past a question were coded as “not applicable” and retained in other analyses. Completely missing responses were rare and handled as missing data.

\subsubsection{Effect of Injury Severity on Perceived Importance of Functional Restoration}
Responses to \textbf{Part 3} (questions 9, 13, and 18) of the survey were used to assess the distribution of participant's perceived importance towards regaining motor and sensory functions. Relative preference (Eq~\ref{eq:relative_preference}) was used to quantify how much more (or less) likely participants with a given characteristic (e.g., prior knowledge of BCI or injury severity) were to express a particular response compared to participants without that characteristic. In most cases, this refers to variations in the level of a specific characteristic, such as differing degrees of prior knowledge about BCI or varying severity of injury. This measure captures the ratio of probabilities between groups, analogous to relative risk in epidemiological studies, and allows us to assess whether certain factors influence preferences or willingness. To handle instances where a cell count was zero (which would otherwise result in undefined or extreme scores), the Haldane-Anscombe correction was applied by adding 0.5 to all cell counts in the 2$\times$2 contingency table. This adjustment provides more stable estimates of the relative preference, particularly in small sample sizes \cite{weber2020zero}. In the case of the participant's perceived importance of functional restoration, the null hypothesis was that the severity of the injury does not change the important/unimportant ratio. A 95\% confidence interval was also calculated using Equation~\ref{eq:confidence_interval_of_relative_risk}.

\begin{figure*}[!htpb]
The equation for relative preference ($RP$):
\begin{equation} \label{eq:relative_preference}
    RP = \frac{a(a+c)^{-1}}{b(b+d)^{-1}} = \frac{a(b+d)}{b(a+c)}
\end{equation}

The equation for standard error ($SE$) of $RP$:
\begin{equation} \label{eq:standard_error_of_relative_risk}
    SE(\log(RP)) = \sqrt{\frac{c}{a(a+c)} + \frac{d}{b(b+d)}}
\end{equation}

The equation for the $1-\alpha$ confidence interval ($CI$) of $RP$:
\begin{equation} \label{eq:confidence_interval_of_relative_risk}
    CI_{1-\alpha}(\log(RP)) = \log(RP) + SE(\log(RP)) \times z_{\alpha}
\end{equation}

where $a$, $b$, $c$, and $d$ are defined in Table~\ref{tab:relative_preference_variable_definition} and $z_\alpha$ is defiend as the standard score for level of significance $\alpha$

\end{figure*}

\begin{table}[h]
\caption{Variable definitions for relative preference calculation (Eq.~\ref{eq:relative_preference},~\ref{eq:standard_error_of_relative_risk})}
\label{tab:relative_preference_variable_definition}
\begin{tabular}{@{}llll@{}}
\toprule
& & \textbf{More Severe} & \textbf{Less Severe} \\
\textbf{} & & Rankin 4, 5 & Rankin 2, 3 \\
\textbf{} & & ASIA A, B & ASIA C, D \\
\midrule
\textbf{More Willing} & Very/Moderately Likely & a & b \\
\textbf{Less Willing} & Slightly/Not at All Likely & c & d \\
\botrule
\end{tabular}
\end{table}

\subsubsection{Effect of Levels of Restoration on Willingness to Undergo ECoG-BCI Implantation}
Responses to  \textbf{Part 3} (questions 10--12, 14--17, 19--23) of the survey were used to determine whether offering more functional capabilities in the ECoG-based BCI devices would lead to an increase in participants’ willingness to undergo surgery to implant the system for upper-extremity motor, lower extremity motor, and sensation functions. Relative preferences to undergo surgery with respect to injury severity to restore the aforementioned functions were calculated using Equation~\ref{eq:relative_preference}. The null hypothesis was that the severity of the injury does not change the willing/unwilling ratio. A 95\% confidence interval was also calculated.

\subsubsection{Effect of Perceived Importance of Functional Restoration on Willingness to Undergo ECoG-BCI Implantation}
Responses to \textbf{Part 3} (questions 9--23) of the survey were used to determine whether an increase in the perceived importance to regain upper-extremity motor, lower extremity motor, and sensation functions would also increase willingness to undergo surgery for the appropriate restorations. Relative preference to undergo surgery with respect to perceived importance of the aforementioned functions were calculated using Equation~\ref{eq:relative_preference}. The null hypothesis was that the perceived importance of a given function does not change the willing/unwilling ratio. A 95\% confidence interval was also calculated using Equation~\ref{eq:confidence_interval_of_relative_risk}.

\subsubsection{Effect of Prior Knowledge on Willingness to Undergo ECoG-BCI Implantation}
Responses to \textbf{Part 2} (question 8) of the survey combined with \textbf{Part 3} (questions 12, 17, 19--21, and 23) was used to determine whether prior ECoG BCI knowledge influenced the relative preference to undergo surgery to restore upper/lower extremity motor or sensation functions. Relative preference (Eq~\ref{eq:relative_preference}) is used to assess whether prior knowledge of implantable BCI influences a participant’s willingness to undergo surgery. The null hypothesis is that given prior knowledge of BCI, the participant is equally as willing to undergo surgery for functional restoration compared to someone without BCI knowledge. A 95\% confidence interval was also calculated using Equation~\ref{eq:confidence_interval_of_relative_risk}.

\subsubsection{Additional Functional Restoration}
Responses to \textbf{Part 4} (questions 24–26) of the survey were analyzed to assess participants’ interest in restoring additional functions beyond those listed in earlier sections, as well as their willingness to undergo surgery for such restorations. Question 25 included an open-ended text box in which participants could freely describe any additional functions they wished to regain. Responses were reviewed and categorized into thematic groups.

\subsubsection{Concerns}
Responses to \textbf{Part 4} (question 28) of the survey was analyzed to identify participants' concerns about undergoing a prospective ECoG-based BCI implantation. The number of participants expressing a specific concern was normalized and presented as a percentage of the total survey participants.

\section{Results}

\begin{table}[htbp!]
\caption{Characteristics of Survey Participants}
\label{tab:data_summary}
\tiny\rm
\begin{tabular}{@{}*{6}{l}}
\toprule
\textbf{Category} & \textbf{Group} & \textbf{Total (\%)} & \textbf{SCI (\%)} & \textbf{Stroke (\%)} & \textbf{Both (\%)} \\ 
\midrule
\textbf{Type of Injury}
& All participants & 71 & 33  & 37  & 1  \\
& SCI              & 34  (47.89) & 33 (100.00) &  -          & 1 (100.00) \\
& Stroke           & 38  (53.52) &  -          & 37 (100.00) & 1 (100.00) \\
\midrule
\textbf{Type of SCI Deficit}
& Tetraplegia                  & 12 (16.90) & 11 (33.33) & - & 1 (100.00) \\
& Paraplegia                   & 15 (21.13) & 15 (45.45) & - & 0   (0.00) \\
& Don't know/Decline to answer &  1  (1.41) &  1  (3.03) & - & 0   (0.00) \\
& Not applicable               &  6  (8.45) &  6 (18.18) & - & 0   (0.00) \\
\midrule
\textbf{Level of SCI Impairment}
& Cervical                     & 17 (23.94) & 16 (48.48) & - & 1 (100.00) \\
& Thoracic                     &  3  (4.23) &  3  (9.09) & - & 0   (0.00) \\
& Lumbar                       &  3  (4.23) &  3  (9.09) & - & 0   (0.00) \\
& Sacral                       &  0  (0.00) &  0  (0.00) & - & 0   (0.00) \\
& Don't know/Decline to answer &  5  (7.04) &  5 (15.15) & - & 0   (0.00) \\
& Not applicable               &  6  (8.45) &  6 (18.18) & - & 0   (0.00) \\
\midrule
\textbf{ASIA Scale}
& Grade A                      &  7  (9.87) &  7 (21.21) & - & 0   (0.00) \\
 & Grade B                      & 12 (16.90) & 12 (36.36) & - & 0   (0.00) \\
 & Grade C                      &  7  (9.87) &  6 (18.18) & - & 1 (100.00) \\
 & Grade D                      &  4  (5.63) &  4 (12.12) & - & 0   (0.00) \\
 & Grade E                      &  0  (0.00) &  0  (0.00) & - & 0   (0.00) \\
 & Don't know/Decline to answer &  4  (5.63) &  4 (12.12) & - & 0   (0.00) \\
\midrule
\textbf{Modified Rankin Scale}
& 0                            &  0  (0.00) & - &  0  (0.00) & 0   (0.00) \\
& 1                            &  0  (0.00) & - &  0  (0.00) & 0   (0.00) \\
& 2                            &  3  (4.23) & - &  3  (8.11) & 0   (0.00) \\
& 3                            & 11 (15.49) & - & 11 (29.73) & 0   (0.00) \\
& 4                            & 11 (15.49) & - & 11 (29.73) & 0   (0.00) \\
& 5                            & 12 (16.90) & - & 11 (29.73) & 1 (100.00) \\
& Don't know/Decline to answer &  1  (1.41) & - &  1  (2.70) & 0   (0.00) \\
\midrule
\textbf{Mobility Assistance}
& Walking aid                  & 11 (15.49) &  3  (9.09) &  8 (21.62) & 0   (0.00) \\
& Wheelchair                   & 15 (21.13) & 11 (33.33) &  3  (8.11) & 1 (100.00) \\
& Other                        &  1  (1.41) &  1  (3.03) &  0  (0.00) & 0   (0.00) \\
& Don't know/Decline to answer &  0  (0.00) &  0  (0.00) &  0  (0.00) & 0   (0.00) \\
& Not applicable               & 44 (61.97) & 18 (54.55) & 26 (70.27) & 0   (0.00) \\
\midrule
\textbf{Age}
& 18-24                        &  9 (12.68) & 9 (27.27) &  0  (0.00) & 0   (0.00) \\
& 25-34                        & 11 (15.49) & 9 (27.27) &  1  (2.70) & 1 (100.00) \\
& 35-44                        &  5  (7.04) & 3  (9.09) &  2  (5.41) & 0   (0.00) \\
& 45-54                        &  7  (9.86) & 2  (6.06) &  5 (13.51) & 0   (0.00) \\
& 55-64                        & 18 (25.35) & 4 (12.12) & 14 (37.84) & 0   (0.00) \\
& 65-74                        & 14 (19.72) & 3  (9.09) & 11 (29.73) & 0   (0.00) \\
& 75 or older                  &  4  (5.63) & 1  (3.03) &  3  (8.11) & 0   (0.00) \\
& Don't know/Decline to answer &  1  (1.41) & 1  (3.03) &  0  (0.00) & 0   (0.00) \\
& Not applicable               &  2  (2.82) & 1  (3.03) &  1  (2.70) & 0   (0.00) \\
\midrule
\textbf{Gender}
& Male              & 48 (67.61) & 22 (66.67) & 25 (67.57) & 1 (100.00) \\
& Female            & 18 (25.35) &  8 (24.24) & 10 (27.03) & 0   (0.00) \\
& Other             &  1  (1.41) &  1  (3.03) &  0  (0.00) & 0   (0.00) \\
& Decline to answer &  1  (1.41) &  0  (0.00) &  1  (2.70) & 0   (0.00) \\
& Not applicable    &  3  (4.23) &  2  (6.06) &  1  (2.70) & 0   (0.00) \\
\midrule
\textbf{Level of Education} 
& Some high school or less                 &  4  (5.63) &  3  (9.09) &  1  (2.70) & 0   (0.00) \\
& High school diploma or equivalent        & 11 (15.49) &  5 (15.15) &  6 (16.22) & 0   (0.00) \\
& Some college or a two year degree        & 21 (29.58) & 10 (30.30) & 11 (29.73) & 0   (0.00) \\
& College graduate                         & 18 (25.35) &  8 (24.24) &  9 (24.32) & 1 (100.00) \\
& Advanced degree (Master's, JD, MD, etc.) & 14 (19.72) &  6 (18.18) &  8 (21.62) & 0   (0.00) \\
& Don't know/Decline to answer             &  1  (1.41) &  0  (0.00) &  1  (2.70) & 0   (0.00) \\
& Not applicable                           &  2  (2.82) &  1  (3.03) &  1  (2.70) & 0   (0.00) \\
\midrule
\textbf{Living situation}
& Nursing facility + Full time assistance &  0  (0.00) &  0  (0.00) &  0  (0.00) & 0   (0.00) \\
& Nursing facility + Partial assistance   &  2  (2.82) &  0  (0.00) &  2  (5.41) & 0   (0.00) \\
& Home + supportive living service        &  6  (8.45) &  4 (12.12) &  2  (5.41) & 0   (0.00) \\
& Home + somebody else                    & 34 (47.89) & 15 (45.45) & 18 (48.65) & 1 (100.00) \\
& Home + no assistance                    & 21 (29.58) &  7 (21.21) & 14 (37.84) & 0   (0.00) \\
& Don't know/Decline to answer            &  0  (0.00) &  0  (0.00) &  0  (0.00) & 0   (0.00) \\
& Not applicable                          &  8 (11.27) &  7 (21.21) &  1  (2.70) & 0   (0.00) \\
\midrule
\textbf{Annual Income}
& \$9,999 or less              &  9 (12.68) &  4 (12.12) & 4 (10.81) & 1 (100.00) \\
& \$10,000-\$19,999            &  3  (4.23) &  2  (6.06) & 1  (2.70) & 0   (0.00) \\
& \$20,000-\$39,999            & 11 (15.49) &  5 (15.15) & 6 (16.22) & 0   (0.00) \\
& \$40,000-\$59,999            & 12 (16.90) &  4 (12.12) & 8 (21.62) & 0   (0.00) \\
& \$60,000-\$69,999            &  5  (7.04) &  2  (6.06) & 3  (8.11) & 0   (0.00) \\
& \$70,000-\$99,999            &  2  (2.82) &  1  (3.03) & 1  (2.70) & 0   (0.00) \\
& \$100,000-\$199,999          & 11 (15.49) &  6 (18.18) & 5 (13.51) & 0   (0.00) \\
& \$200,000 or higher          &  1  (1.41) &  1  (3.03) & 0  (0.00) & 0   (0.00) \\
& Don't know/Decline to answer & 15 (21.13) &  7 (21.21) & 8 (21.62) & 0   (0.00) \\ 
& Not applicable               &  2  (2.82) &  1  (3.03) & 1  (2.70) & 0   (0.00) \\
\bottomrule
\end{tabular}
\end{table}

\subsection{Participant Classification}
A total of 71 participants responded to the survey. The participant injury profile and demographic information are summarized in Table~\ref{tab:data_summary}. Of the 71 participants, 4 did not provide sufficient information on the ASIA Impairment Scale and were excluded from subgroup analyses involving injury severity. This resulted in a final analytic sample of 12 with tetraplegia and 15 with paraplegia.

\begin{figure*}[!htbp]
    \centering
    \includegraphics[width=5in]{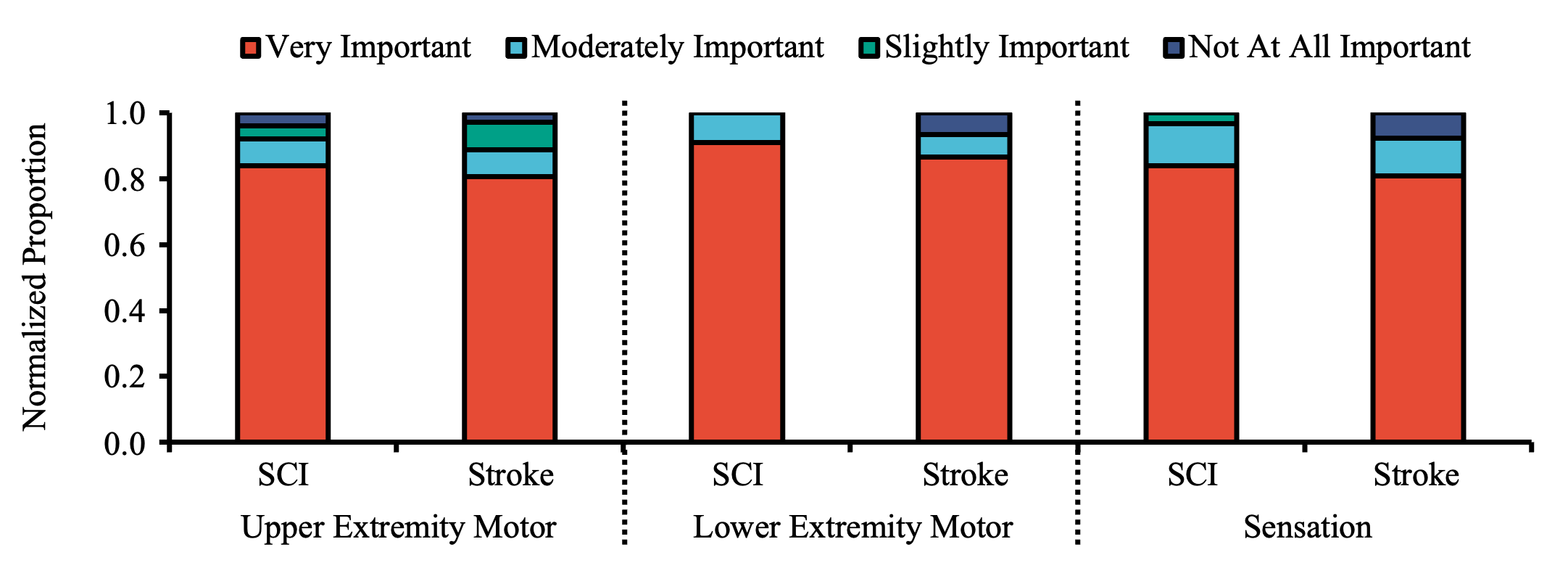}
    \caption{Importance of regaining motor and sensory functions to stroke and SCI participants as a normalized proportion, excluding ``Does Not Apply" and ``Don't Know/Decline" (included in Fig~\ref{fig:participant_perception_sup}). The majority ($>80\%$) of both stroke and SCI participants indicated that regaining upper extremity motor function, lower extremity motor function, and sensation were all ``Very Important." }
    \label{fig:participant_perception}
\end{figure*}

\subsection{Effect of Injury Severity on Perceived Importance of Functional Restoration}
Overall, SCI participants perceived lower-extremity motor function to be more important than upper-extremity motor function and sensation/bladder functions; Stroke participants perceived both upper- and lower-extremity motor functions to be slightly more important than sensation/bladder functions (Fig~\ref{fig:participant_perception}). Additionally, a larger proportion of SCI participants considered sensory function to be very important compared to stroke participants. 

\begin{figure*}[!htbp]
    \centering
    \includegraphics[width=5in]{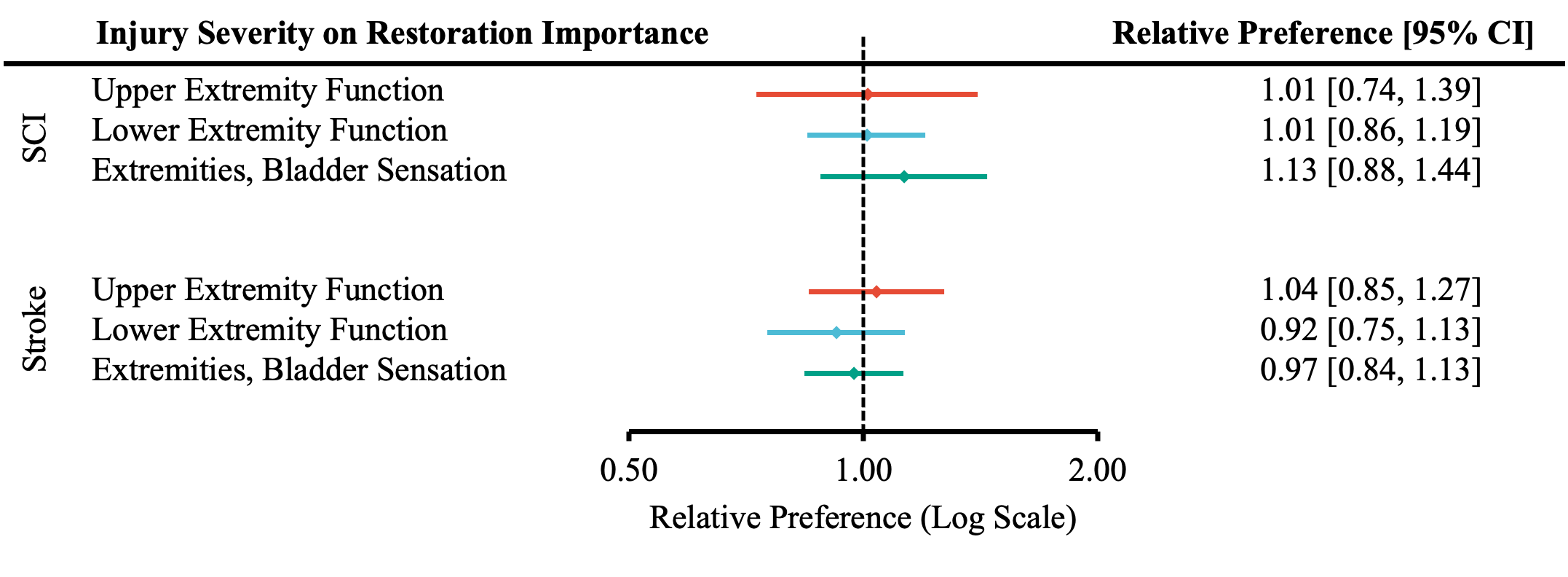}
    \caption{Forest plot of the relative preferences on whether injury severity influenced participants' perceived importance of functional restoration. According to Table~\ref{tab:relative_preference_variable_definition}, the number of participants indicated \textit{Very Important} and \textit{Moderately Important} is placed in the ``More Willing'' group; \textit{Slight Important} and \textit{Not at all Important} in the ``Less Willing'' group; \textit{ASIA A, B} or \textit{Rankin 4, 5} in the ``More Severe'' group; \textit{ASIA C, D} or \textit{Rankin 2, 3} in the ``Less Severe'' group.}
    \label{fig:forest_importance_severity}
\end{figure*}

The severity of disability did not significantly influence the perceived importance of restoring upper extremity motor, lower extremity motor, or sensory function (no statistically significant differences observed at the 95\% confidence level, Fig~\ref{fig:forest_importance_severity}). 

\begin{figure*}[!htbp]
    \centering
    \includegraphics[width=5in]{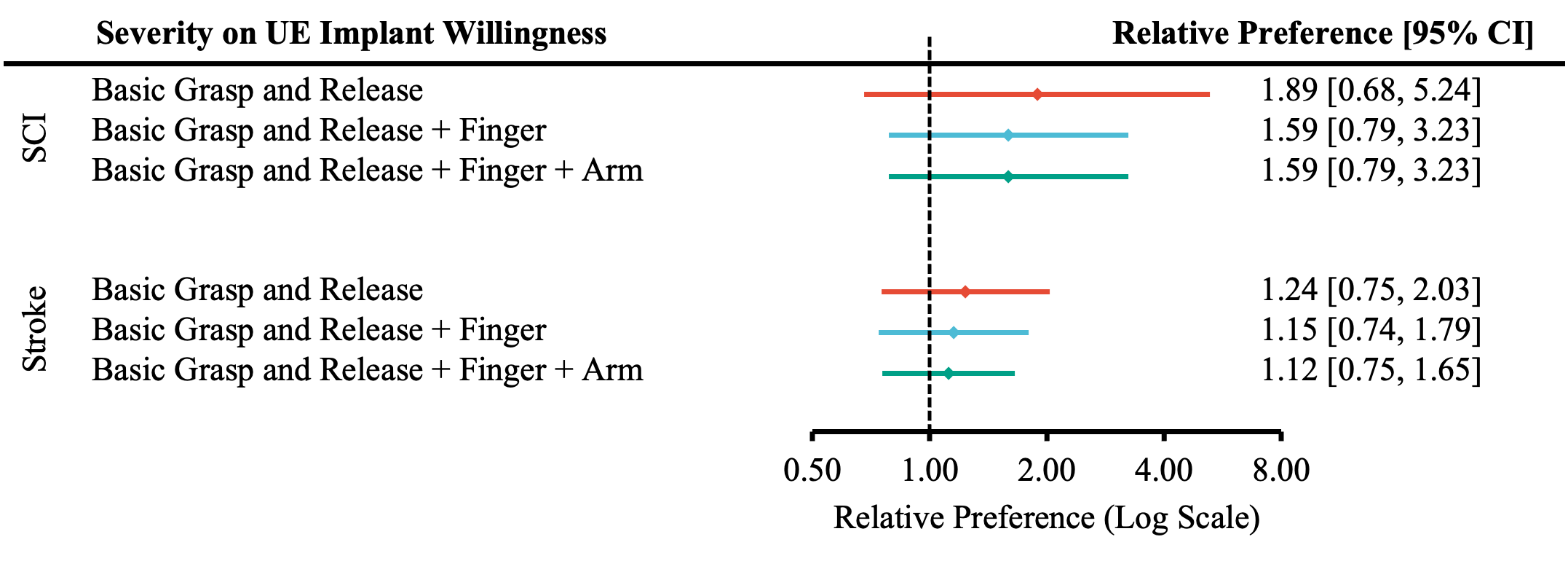}
    \caption{Forest plot of the relative preference to undergo surgery to restore upper-extremity functions with respect to injury severity. According to Table~\ref{tab:relative_preference_variable_definition}, the number of participants indicated \textit{Very Likely} and \textit{Moderately Likely} is placed in the ``More Willing'' group; \textit{Slight Likely} and \textit{Not at all Likely} in the ``Less Willing'' group; \textit{ASIA A, B} or \textit{Rankin 4, 5} in the ``More Severe'' group; \textit{ASIA C, D} or \textit{Rankin 2, 3} in the ``Less Severe'' group.}
    \label{fig:forest_upper_severity_willingness}
\end{figure*}

\begin{figure*}[!htbp]
    \centering
    \includegraphics[width=5in]{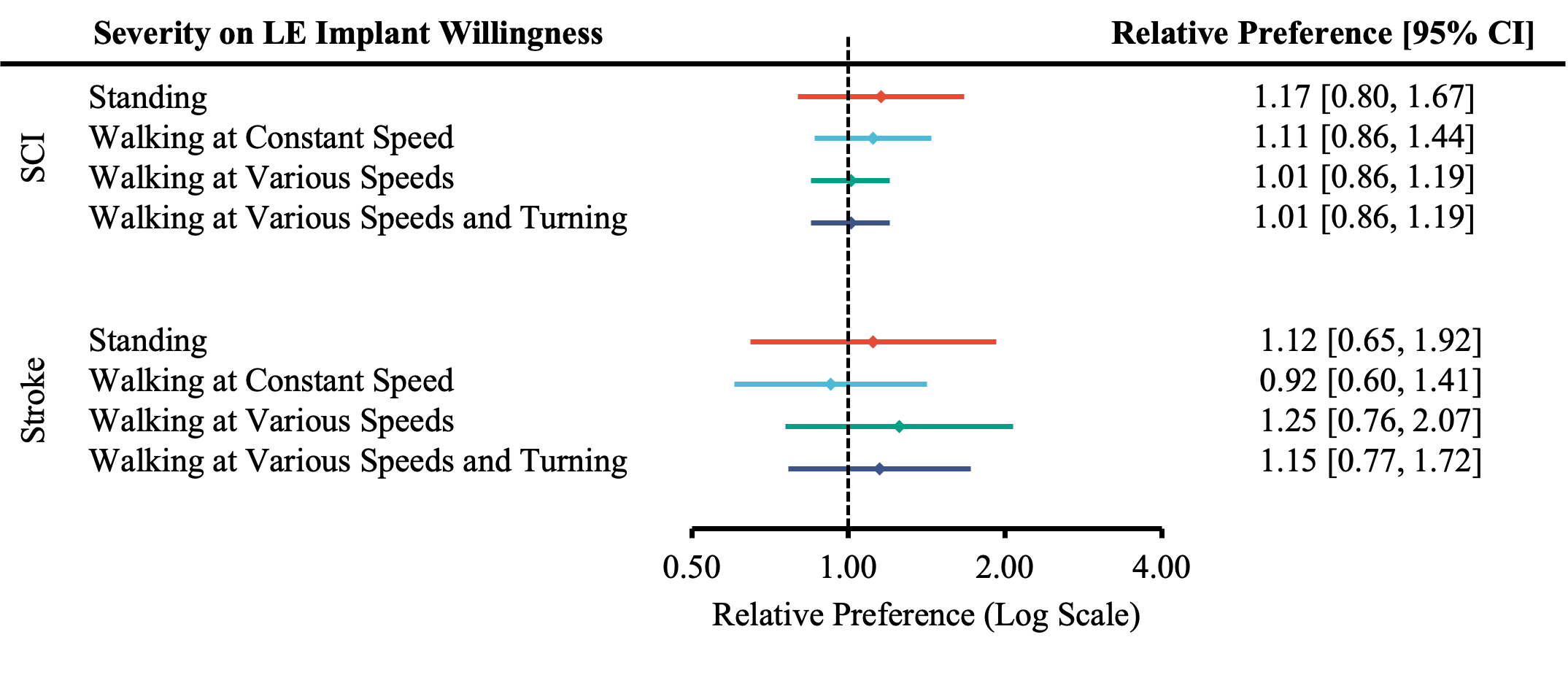}
    \caption{Forest plot of the relative preference to undergo surgery to restore lower-extremity functions with respect to injury severity. According to Table~\ref{tab:relative_preference_variable_definition}, the number of participants indicated \textit{Very Likely} and \textit{Moderately Likely} is placed in the ``More Willing'' group; \textit{Slight Likely} and \textit{Not at all Likely} in the ``Less Willing'' group; \textit{ASIA A, B} or \textit{Rankin 4, 5} in the ``More Severe'' group; \textit{ASIA C, D} or \textit{Rankin 2, 3} in the ``Less Severe'' group.}
    \label{fig:forest_lower_severity_willingness}
\end{figure*}

\begin{figure*}[!htbp]
    \centering
    \includegraphics[width=5in]{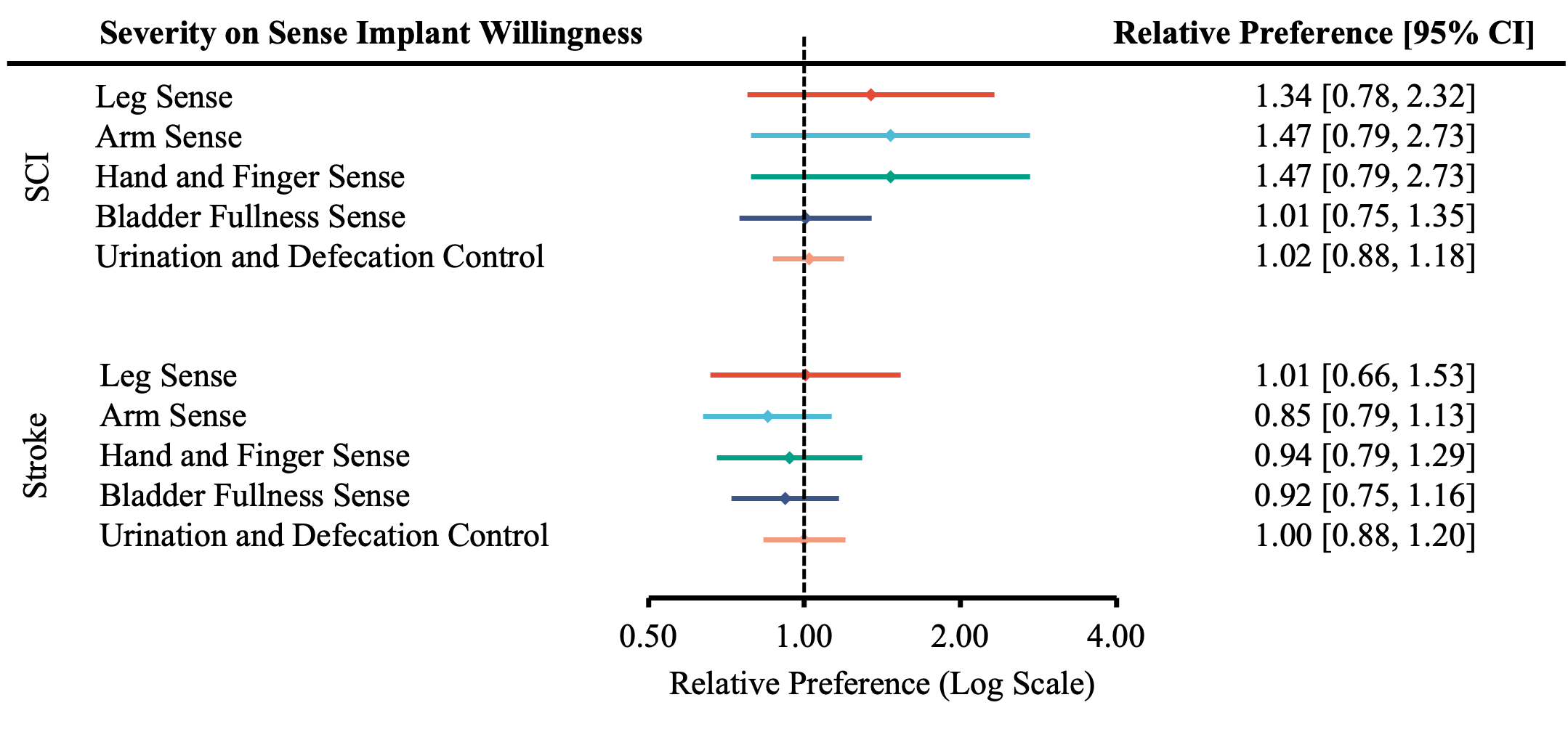}
    \caption{Forest plot of the relative preference to undergo surgery to restore sensory and autonomic functions with respect to injury severity. According to Table~\ref{tab:relative_preference_variable_definition}, the number of participants indicated \textit{Very Likely} and \textit{Moderately Likely} is placed in the ``More Willing'' group; \textit{Slight Likely} and \textit{Not at all Likely} in the ``Less Willing'' group; \textit{ASIA A, B} or \textit{Rankin 4, 5} in the ``More Severe'' group; \textit{ASIA C, D} or \textit{Rankin 2, 3} in the ``Less Severe'' group.}
    \label{fig:forest_sensory_severity_willingness}
\end{figure*}

The relative preference for willingness to undergo surgery based on the degree of injury was not statistically significant (Fig~\ref{fig:forest_upper_severity_willingness},\ref{fig:forest_lower_severity_willingness},\ref{fig:forest_sensory_severity_willingness}).

\begin{figure*}[!htbp]
    \centering
    \includegraphics[width=5in]{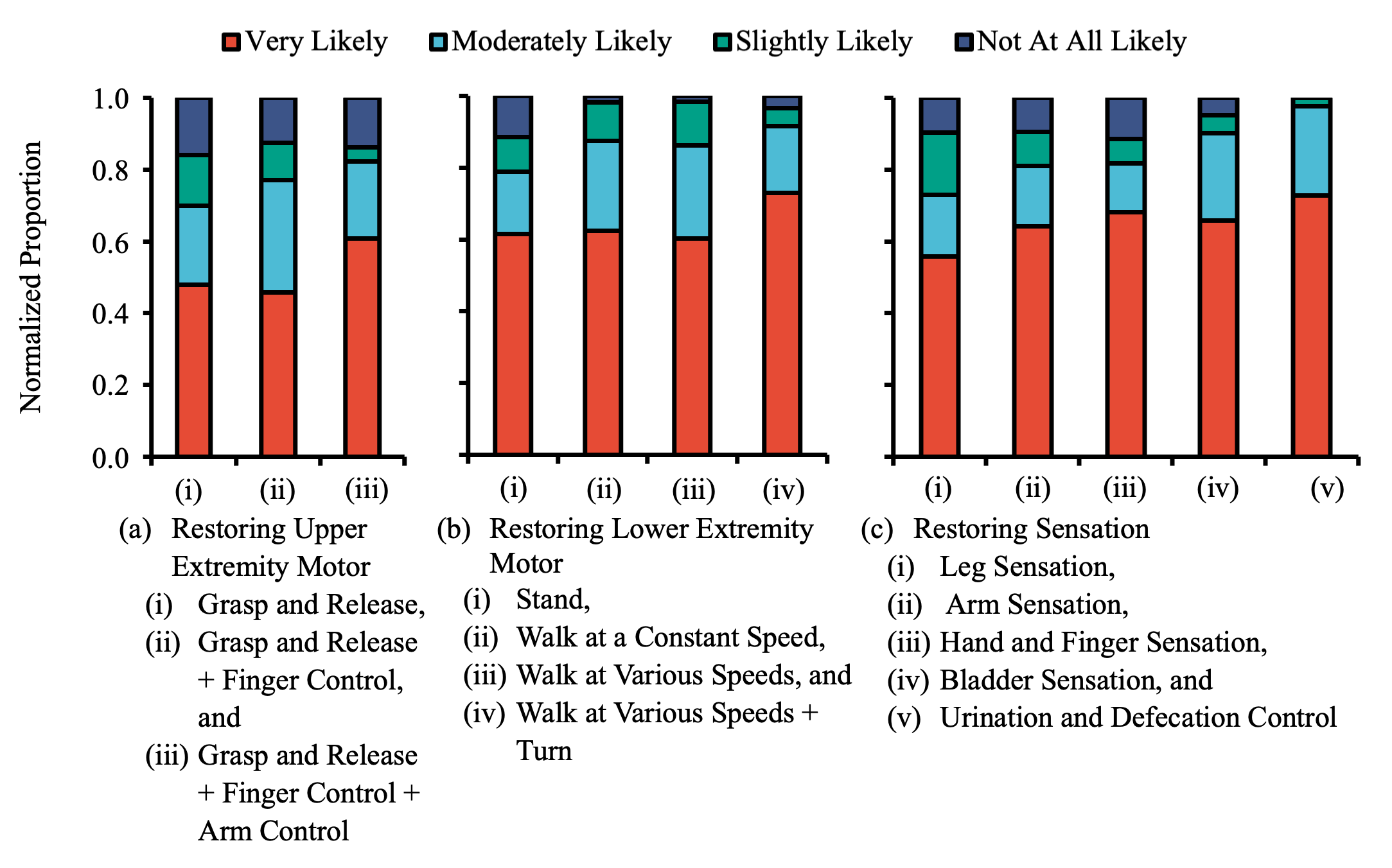}
    \caption{Participant willingness to undergo surgery to implant BCI system at various degrees of function restored. (a) Upper extremity motor function. (b) Lower extremity motor function. (c) Sensation. Participants were asked to indicate their level of disability as well as how important regaining upper extremity motor function, lower extremity motor function, and sensation were to them. Participants’ responses were plotted as a sample size-normalized stacked bar graph of those who indicated that each function was ``very important,'' ``moderately important,'' ``slightly important,'' ``not at all important,'' or ``does not apply.'' The majority of participants were at least moderately willing to undergo surgery to implant BCI systems for even basic levels of functional restoration.}
    \label{fig:willingness_to_undergo_surgery}
\end{figure*}

\subsection{Willingness to Undergo ECoG-BCI Implantation}
The majority of participants responded ``Very likely'' or ``Moderately likely'' when asked about their willingness to undergo surgery for an ECoG-based BCI to restore upper extremity motor, lower extremity motor, sensation, and bladder functions (Fig~\ref{fig:willingness_to_undergo_surgery}). Specifically, for upper extremity function, 70\% of participants stated they would be ``Very likely'' or ``Moderately likely'' to undergo surgery to restore ``Grasp and Release,'' 77\% for ``Grasp and Release + Fine Finger Control,'' and 82\% for ``Grasp and Release + Fine Finger Control + Fine Arm Control.'' For lower extremity function, 79\% of participants stated they would be ``Very likely'' or ``Moderately likely'' to undergo surgery to restore ``Stand,'' 88\% for ``Walk at a Constant Speed,'' 86\% for ``Walk at Various Speeds,'' and 92\% for ``Walk at Various Speeds + Turn.'' For sensation and urination/defecation control 73\% of participants stated they would be ``Very likely'' or ``Moderately likely'' to undergo surgery to restore ``Leg Sensation,'' 81\% for ``Arm Sensation,'' 82\% for ``Hand and Finger Sensation,'' 90\% for ``Bladder Sensation,'' and 98\% for ``Urination and Defecation Control.''

\subsection{Effect of Levels of Restoration on Willingness to Undergo ECoG-BCI Implantation}
Responses did not significantly differ based on the degree of functionality restored (Fig~\ref{fig:willingness_to_undergo_surgery} illustrates this for upper extremity motor functions).

\begin{figure*}[!htbp]
    \centering
    \includegraphics[width=5in]{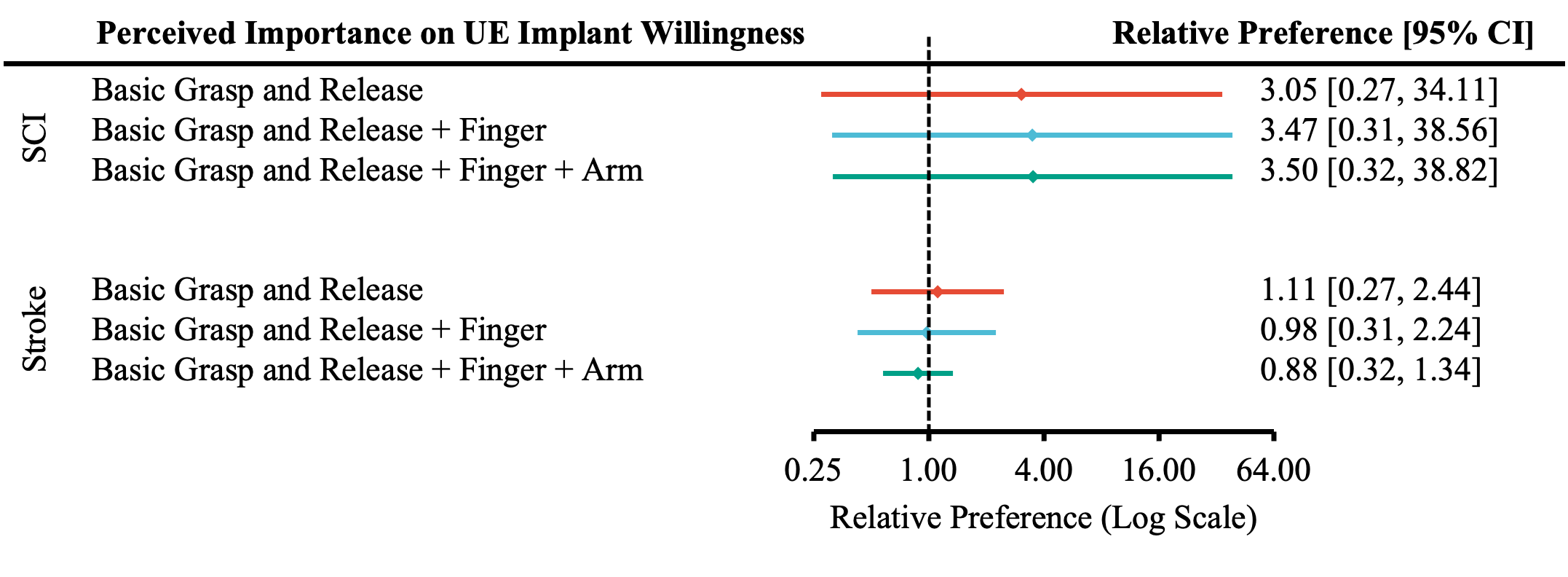}
    \caption{Forest plot of the relative preference to undergo surgery to restore upper-extremity functions with respect to perceived importance of functional restoration. According to Table~\ref{tab:relative_preference_variable_definition}, the number of participants indicated \textit{Very Likely} and \textit{Moderately Likely} is placed in the ``More Willing'' group; \textit{Slight Likely} and \textit{Not at all Likely} in the ``Less Willing'' group; \textit{Very Important} and \textit{Moderately Important} in the ``More Important'' group; \textit{Slight Important} and \textit{Not at all Important} in the ``Less Important'' group.}
    \label{fig:forest_upper_importance_willingness}
\end{figure*}

\begin{figure*}[!htbp]
    \centering
    \includegraphics[width=5in]{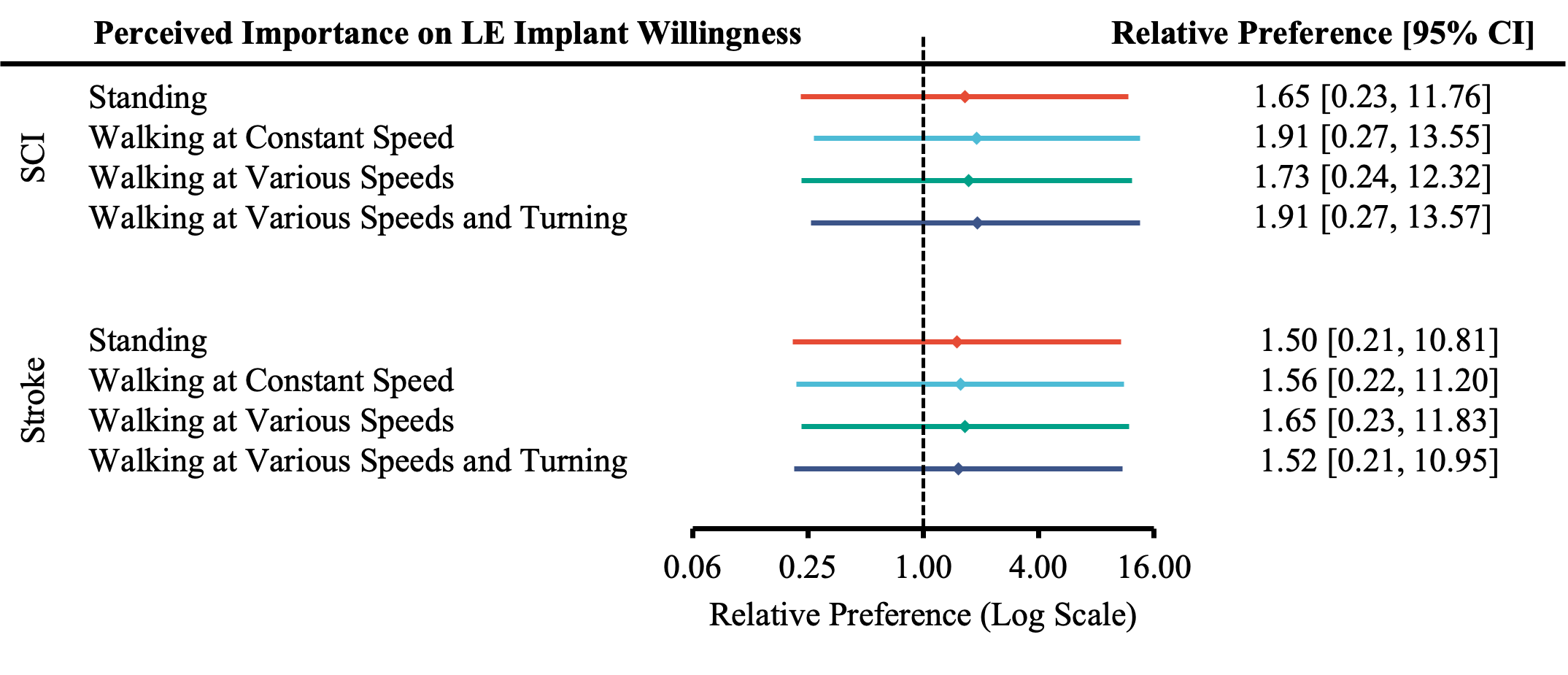}
    \caption{Forest plot of the relative preference to undergo surgery to restore lower-extremity functions with respect to perceived importance of functional restoration. According to Table~\ref{tab:relative_preference_variable_definition}, the number of participants indicated \textit{Very Likely} and \textit{Moderately Likely} is placed in the ``More Willing'' group; \textit{Slight Likely} and \textit{Not at all Likely} in the ``Less Willing'' group; \textit{Very Important} and \textit{Moderately Important} in the ``More Important'' group; \textit{Slight Important} and \textit{Not at all Important} in the ``Less Important'' group.}
    \label{fig:forest_lower_importance_willingness}
\end{figure*}

\begin{figure*}[!htbp]
    \centering
    \includegraphics[width=5in]{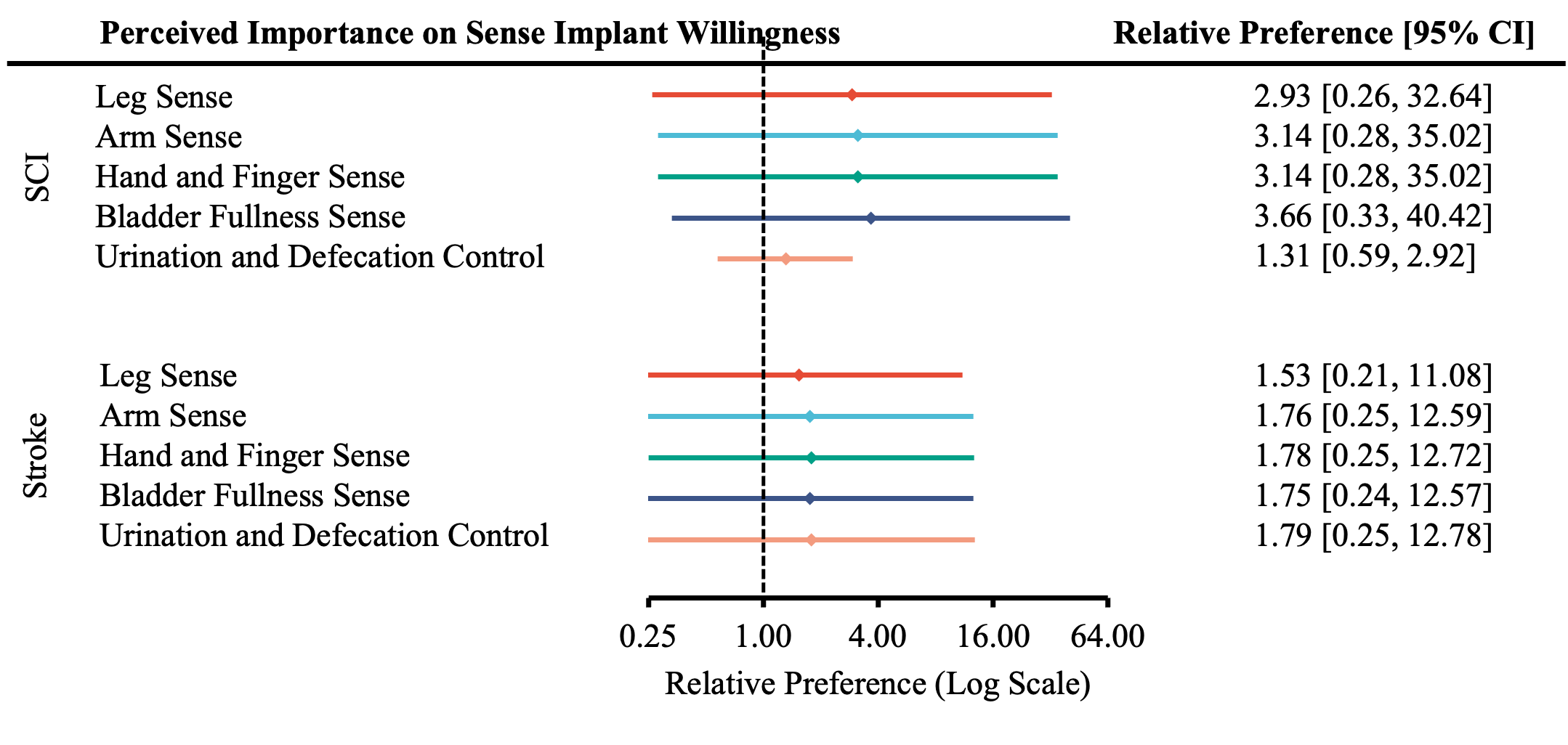}
    \caption{Forest plot of the relative preference to undergo surgery to restore sensory and autonomic functions with respect to perceived importance of functional restoration. According to Table~\ref{tab:relative_preference_variable_definition}, the number of participants indicated \textit{Very Likely} and \textit{Moderately Likely} is placed in the ``More Willing'' group; \textit{Slight Likely} and \textit{Not at all Likely} in the ``Less Willing'' group; \textit{Very Important} and \textit{Moderately Important} in the ``More Important'' group; \textit{Slight Important} and \textit{Not at all Important} in the ``Less Important'' group.}
    \label{fig:forest_sensory_importance_willingness}
\end{figure*}

\subsection{Effect of Perceived Importance of Functional Restoration on Willingness to Undergo Surgery}
In all levels of functional restoration, most applicable participants reported ``Very Likely'' to undergo surgery and rated function restoration as ``Very Important'' (Fig~\ref{fig:sci_restoration_vs_sx},~\ref{fig:stroke_restoration_vs_sx}). The effect of perceived importance of functional restoration on the relative preference for willingness to undergo surgery was not statistically significant in any case (Fig~\ref{fig:forest_upper_importance_willingness},\ref{fig:forest_lower_importance_willingness},\ref{fig:forest_sensory_importance_willingness}).

\begin{figure*}[!htbp]
    \centering
    \includegraphics[width=5in]{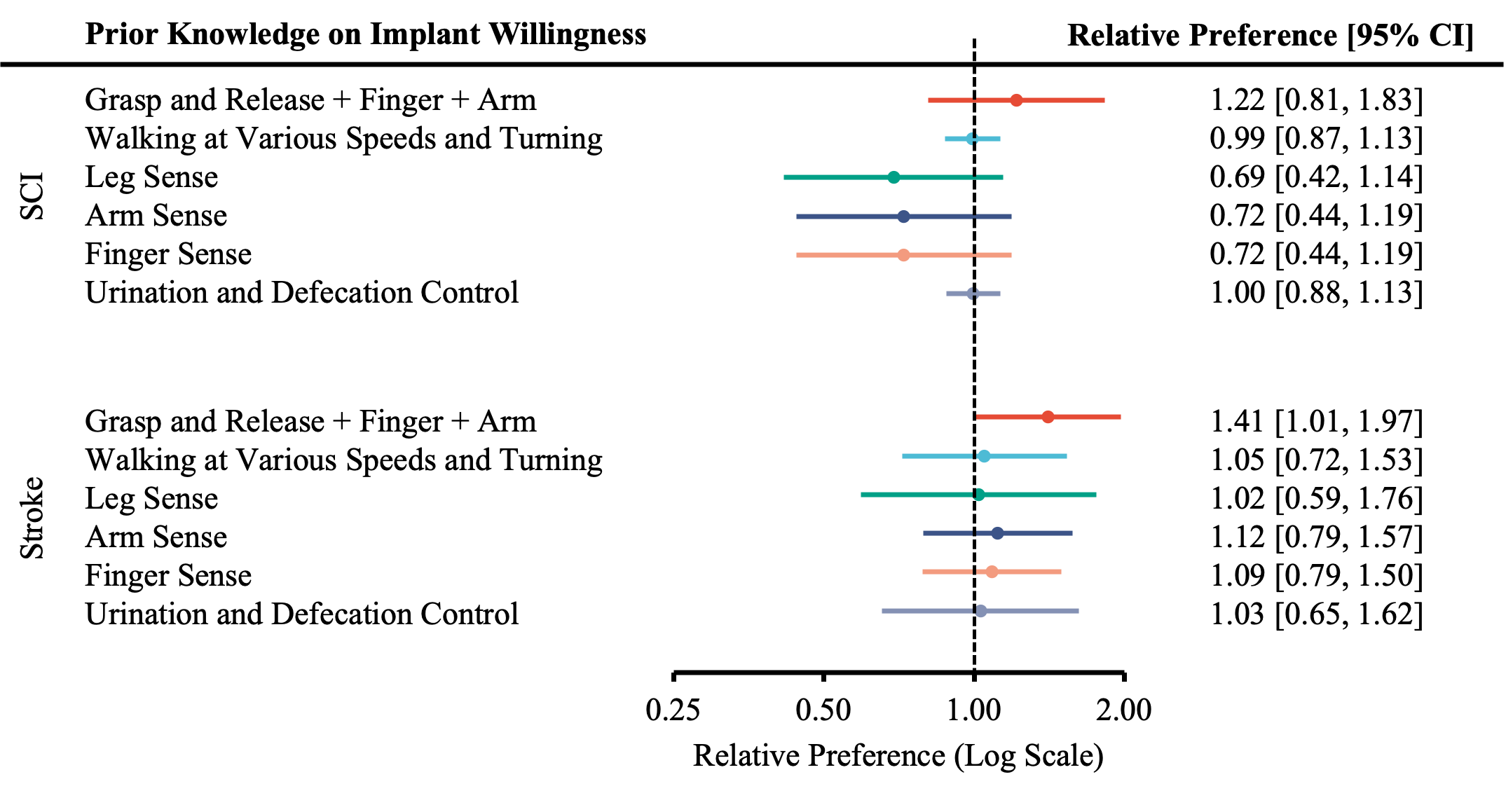}
    \caption{Relative willingness of participants to undergo surgery for device implantation to restore a particular function with or without prior knowledge. The number of participants with prior BCI knowledge is placed in the intervention group; without prior knowledge in the control group; participants that are more willing to undergo surgery in the event group; less willing in the non-event group.}
    \label{fig:prior_knowledge}
\end{figure*}

\subsection{Effect of Prior Knowledge on Willingness to Undergo ECoG-BCI Implantation}
Prior knowledge of BCI technology did not influence willingness to undergo surgery to restore functions with one's exception (Fig~\ref{fig:prior_knowledge}). Specifically, stroke participants with prior knowledge of BCI technology expressed more willingness to undergo surgical implantation of ECoG-based BCIs to restore a combination of arm, grasp/release, and finger control function.

\begin{figure*}[htbp]
    \centering
    \includegraphics[width=5in]{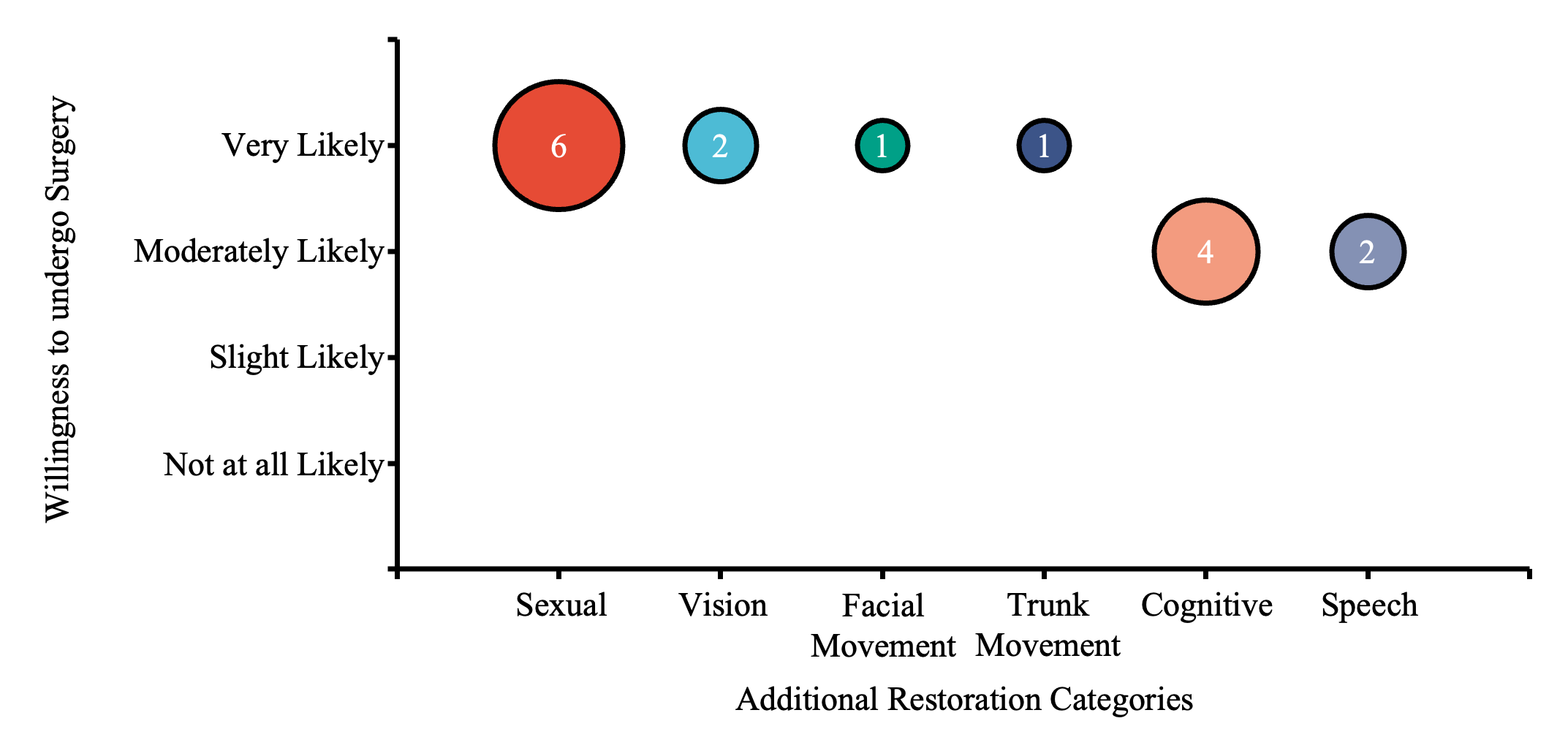}
    \caption{A bubble chart illustrating willingness to undergo surgery for additional restoration categories. The bubble size represents the number of participants interested in each category, and the categories are ordered by average likelihood of undergoing surgery. The number in the bubble represents the number of answers. Willingness was rated on a scale from “Not at all Likely” to “Very Likely,” with an average likelihood calculated for each category. Sexual restoration had the highest interest, followed by Vision, Facial Movement, Trunk Movement, Cognitive, and Speech.}
    \label{fig:additional_restoration}
\end{figure*}

\subsection{Additional Functional Restoration}
In Part 4, the additional restorations suggested by participants were categorized and ranked based on their average willingness to undergo surgery for a given function (Fig~\ref{fig:additional_restoration}). The most frequently endorsed category was Sexual function, with six participants indicating they were “Very Likely” on average to undergo surgery for its restoration. Other notable categories in descending order of importance are vision, facial Movement, trunk movement, cognitive function, and speech.

\begin{figure}[!htpb]
    \centering
    \includegraphics[width=2.5in]{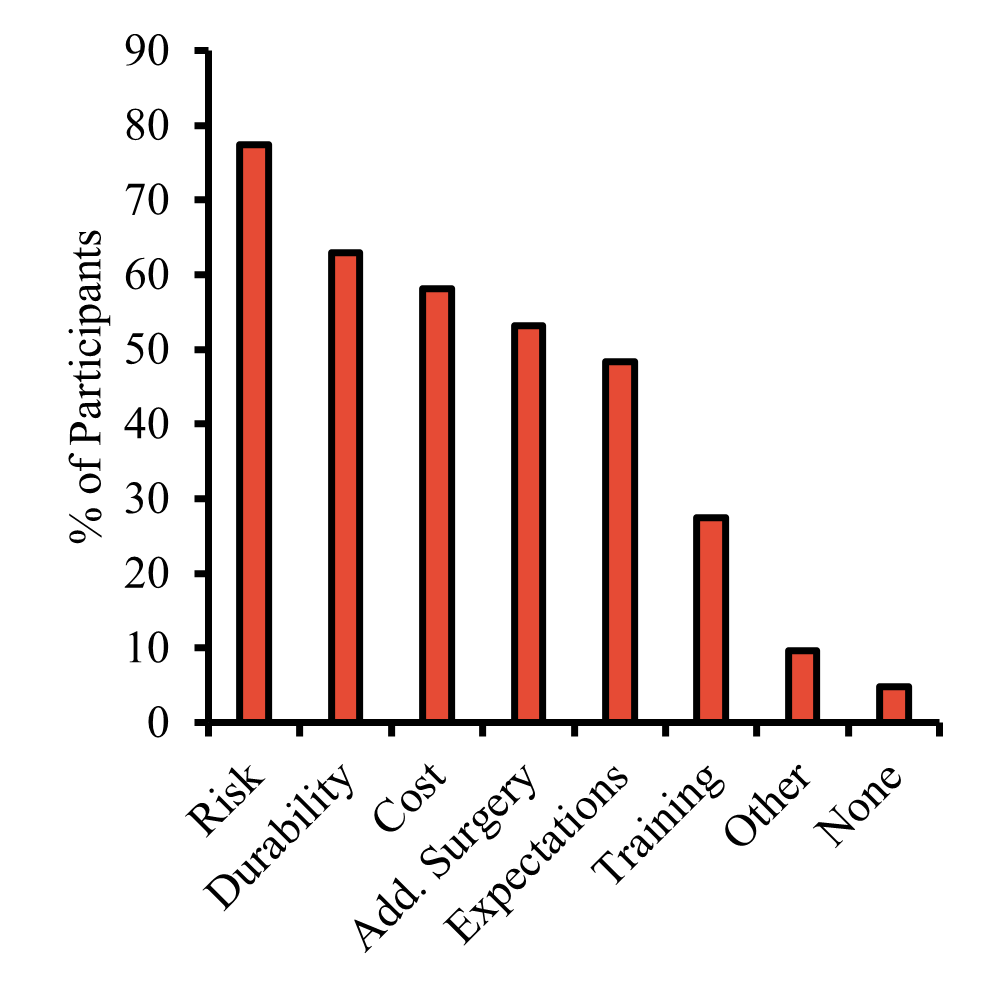}
    \caption{Rank ordered categories of concerns regarding undergoing prospective ECoG-based BCI implantation. Participants were highly concerned about surgical risk and potential complications. Other concerns included cost, the need for additional surgeries, failure of implanted BCI to meet practical expectations, time needed for BCI to be trained, etc.}
    \label{fig:concerns}
\end{figure}

\subsection{Concerns}
In part 4, broad categories of concerns raised by the participants are rank ordered and summarized in Fig~\ref{fig:concerns}. Potential risks and possible complications of surgery, such as infection, scarring, excessive bleeding, blood clots, and reactions to anesthesia, were the most commonly cited concern (77\%). Next, 62\% raised concerns about long-term usage and durability of device. 

\section{Discussion}

This study examined the receptiveness of stroke and SCI cohorts to undergoing surgery to implant ECoG-based BCIs and how it may have been influenced by rehabilitation priorities and prospective level of functional recovery. The results indicate that the stroke population may focus more on upper extremity restoration over lower extremity restoration, and vice versa for the SCI population. The vast majority of stroke and SCI participants were very or moderately willing to undergo surgical implantation of an ECoG-based BCI to restore a given upper/lower extremity motor/sensory as well as autonomic functions. Willingness to undergo surgery was not influenced by the promise of more restorative features within the motor and sensory domains. Disability severity also did not affect willingness to undergo an implantation of a BCI device. Similarly, willingness to undergo surgery was not affected by prior knowledge of BCI technology, with the only exception that stroke participants with prior BCI knowledge were more willing to undergo surgical BCI implantation if it could restore grasp/release finger and arm control. Participants indicated that major concerns for such a surgery included potential surgical risks (e.g., infection, bleeding, and anesthesia complications), the need for additional surgeries if the BCI fails, concerns about long-term durability and reliability of the device, high cost, extensive training time, and the possibility that the BCI may not meet functional expectations.

Functional priorities found in this cohort are consistent with prior literature reports on rehabilitation priorities in the SCI and stroke population \cite{lam2022functional, lo2016functional}. Specifically, the stroke survivor population likely places more emphasis on upper extremities given that many (70\%) are chronically affected by significant upper extremity impairment \cite{dutta2022prevalence}, whereas  $\sim$80\% are ambulatory in the chronic phase \cite{eng2007gait, purton2023stroke}. As such, it is not surprising that restoration of upper extremity function receives more attention. Conversely, chronic SCI participants who have long-term disabilities are almost universally affected by gait impairment, with 73.3\% of participants reporting being wheelchair dependent (Table~\ref{tab:data_summary}, excluding ``Not applicable''). This likely explains why functional restoration of the lower extremities is seen as a higher priority in a larger proportion of SCI participants compared to upper extremity or sensation functions. 

It was unexpected that willingness to undergo surgery for implantation of the ECoG grid and BCI device was not clearly associated with the extent of disability, the perceived importance of functional restoration, or prior knowledge of BCIs. Furthermore, contrary to our expectations, participants did not express greater willingness to undergo surgery when more functions could potentially be restored. Instead, there appeared to be a universal enthusiasm for the implantation of ECoG-based BCIs for functional restoration in general. Given the novelty and perceived promise of BCI technology, it is crucial to consider the ethical implications. Patients with severe disabilities may be especially susceptible to optimistic portrayals of emerging neurotechnologies, potentially leading to premature or inadequately informed consent. Clear clinical guidelines and robust consent frameworks will be essential to safeguard against the risk of unnecessary surgical interventions driven by technological enthusiasm rather than established benefit.

A notable exception to this is that prior knowledge played a role in willingness to undergo BCI implantation surgery for upper extremity motor function restoration amongst stroke participants. It is possible that stroke survivors are more likely to actively search for means to restore upper extremity function, and therefore have likely been exposed to BCI. In turn, this may have galvanized them towards acceptance of BCI technology. Nevertheless, the overarching observation suggests to BCI developers that the first implantable BCI products may not have to provide drastic levels of complexity to gain acceptance with potential recipients. Similarly, BCI developers can also expect that the potential market may not necessarily only involve participants with severe disability. For example, although not explicitly studied here, it may be inferred that participants with even moderate disability would still be willing to undergo surgery for BCI systems that could improve neurological functions. This is opposed to simply being used as a neuroprosthetic for participants with complete loss of neurological function.

Whereas it was previously unknown how sensory restoration via BCI technology would be received among the SCI and stroke survivor populations, this study found that interest in the restoration of somatic sensation in the upper and lower extremities is high. This outcome suggests that invasive ECoG-BCI systems that offer bi-directional features may also be of significant interest. 

Amongst those affected by bladder or bowel function and sensation, there was a very high level of interest in implantable ECoG-based BCI systems that can hypothetically restore such autonomic functions. Similar to prior studies \cite{anderson2004targeting, ditunno2008wants}, bladder and bowel functions have often been rated as equal to or higher in priority than upper or lower extremity motor functions, particularly among individuals with high-level spinal cord injuries. With very limited treatment options for neurogenic bladder and bowel following stroke and SCI, these results suggest that BCI systems seeking to address these issues would be well received. In particular, there are no reported BCI systems that attempt to address bowel and bladder functions at the time of this report and this should be considered an area that merits significant research effort.

A potential limitation of this study is participation bias in the surveyed cohort. In particular, this population may have a tendency to actively seek rehabilitation services and social support services and therefore may be more receptive to emerging technological solutions to restore lost functions. However, this again highlights the enthusiasm for advances in the field of motor and sensory restoration amongst those with disabling neurological injuries. Another limitation of this study is that a significant number of participants omitted a response to what type of mobility aide they use (question 7). It is unknown why this occurred (technical or question not written clearly). Nevertheless, the rate of wheelchair reliance in SCI (73.3\%) and stroke (27.3\%) among participants is consistent with that in the literature \cite{mountain2010rates, ekiz2014wheelchair}.

In summary, the majority of both stroke and SCI participants expressed a high willingness to undergo surgery for ECoG-based BCI implantation, regardless of the extent of their disability or the specific level of function offered by the device. This suggests that even relatively simple or narrowly focused implantable BCI systems may find a receptive user base among individuals with neurological injuries. Notably, bi-directional BCIs (BDBCI) that can restore both motor and sensory functions—particularly grasp and somatic sensation—were of substantial interest, especially among stroke survivors with prior BCI knowledge. Additionally, systems targeting autonomic functions, such as bladder and bowel control, generated strong interest across both populations, reaffirming findings from prior rehabilitation surveys and highlighting a critical unmet need. Taken together, these results provide a strong signal to the BCI research community: there is a clear and enthusiastic demand for clinically viable, implantable BCI systems, especially those that go beyond traditional neuroprosthetics to address complex motor, sensory, and autonomic deficits. Researchers and developers seeking impactful directions should highly consider focusing efforts on bi-directional control, restoration of upper/lower extremity function, and innovations in bladder/bowel BCIs, which represent areas where clinical relevance and user interest are strongly aligned.

\section{Declarations}

This study was deemed to be exempt by the University of California, Irvine Institutional Review Board.

All authors have reviewed and agree to the content of the manuscript.

All authors have no competing interests to declare. 

This study was partially supported by the National Science Foundation (NSF) Grant number 1646275 and internal University of California, Irvine funds. \\

\noindent Data can be made available upon request. 

\bibliography{arxiv}

\newpage

\begin{appendices}

\section{Survey}\label{sec:survey}

\begin{center}
Top of Form
\end{center}

\hrule
\vspace{0.25cm}

\begin{center}
University of California, Irvine

Study Information Sheet
\end{center}

\begin{center}
Functional Priorities of Stroke and Spinal Cord Injury Patients for Invasive Procedures
\end{center}

\begin{center}
Lead Researcher

Tracie Tran, Jr. Specialist, UC Irvine BCI Lab

Neurology

traciet@uci.edu
\end{center}

\begin{center}
Faculty Sponsor

An Do, M.D., Assistant Professor

Neurology

and@uci.edu
\end{center}

\begin{center}
Other Researchers

Zoran Nenadic, Ph.D., Assistant Professor

Biomedical Engineering

znenadic@uci.edu
\end{center}

\begin{center}
Gabrielle Matias, Jr. Specialist, UC Irvine BCI Lab

Neurology

matiasje@uci.edu
\end{center}

\begin{itemize}
    \item You are being asked to participate in a research study that is intended to help us understand your priorities and concerns for future research in emerging technologies that are capable of restoring movement and sensation after spinal cord injury or stroke. We hope to use your insights to improve future studies in this field by accommodating your needs and expectations. 
    \item You are eligible to participate in this study if you 18 years of age or older and suffer from motor impairment due to stroke or spinal cord injury. 
    \item The research procedures involve an online anonymous survey that is estimated to take approximately 10 minutes to complete. Upon completion, you will receive a \$10 Amazon gift card as a token of appreciation for your time. 
    \item Possible risks/discomforts associated with the study include a potential breach of confidentiality, although the utmost care will be taken to avoid this. 
    \item There are no direct benefits from participation in the study.  However, this study may yield insights into current concerns and priorities of stroke and spinal cord injury populations that may guide future brain computer interface research. 
    \item All research data collected will be stored securely and confidentially. No personal identifiers will be collected or stored as the survey will remain anonymous.
    \item The research team, authorized UCI personnel, and regulatory entities, may have access to your study records to protect your safety and welfare.  Any information derived from this research project that personally identifies you will not be voluntarily released or disclosed by these entities without your separate consent, except as specifically required by law. 
    \item If you have any comments, concerns, or questions regarding the conduct of this research please contact the researchers listed at the top of this form. 
    \item Please contact UCI’s Office of Research by phone, (949) 824-6662, by e-mail at IRB@research.uci.edu or at 5171 California Avenue, Suite 150, Irvine, CA 92697 if you are unable to reach the researchers listed at the top of the form and have general questions; have concerns or complaints about the research; have questions about your rights as a research subject; or have general comments or suggestions.
    \item Participation in this study is voluntary.  There is no cost to you for participating.  You may choose to skip a question or a study procedure. You may refuse to participate or discontinue your involvement at any time without penalty.  You are free to withdraw from this study at any time. If you decide to withdraw from this study you should notify the research team immediately. 
    \item As this survey is anonymous, the study team may not be able to extract or delete any specific data provided, should the subject choose to withdraw from the study. 
    \item As part of the completion of this survey, you are agreeing to the “Terms of Use” of eSurv.org, the entity administering the survey.  The data you provide may be collected and used by this agent, as per its privacy agreement.  Note: There is no reasonable expectation that data is anonymous.
\end{itemize}

\begin{center}
By clicking “Next,” you are consenting to partake in the survey.
\end{center}

\vspace{0.25cm}
\hrule
\vspace{0.25cm}

\noindent 1. Please select the type of injury you have.
\begin{itemize}[label=$\square$]
    \item Spinal Cord Injury
    \item Stroke
    \item Don't know/Decline to Answer
\end{itemize}

\vspace{0.25cm}
\hrule
\vspace{0.25cm}

\noindent 2. If known, what is your type of injury?
\begin{itemize}[label=$\square$]
    \item Tetraplegia (Unable to move/feel both arms and legs) 
    \item Paraplegia (Unable to move/feel lower half of body) 
    \item I do not know/Decline to Answer 
\end{itemize}

\noindent 3. Please indicate your level of injury on the ASIA Impairment Scale \textit{(if known)}.
\begin{itemize}[label=$\square$]
    \item Grade A: Complete: No sensory or motor function is preserved in the sacral segments S4-S5. 
    \item Grade B: Incomplete: Sensory but not motor function is preserved below the neurological level and includes the sacral segments S4-S5. 
    \item Grade C: Incomplete: Motor function is preserved below the neurological level, and more than half of the key muscles below the neurological level have a muscle grade less than 3. 
    \item Grade D: Incomplete: Motor function is preserved below the neurological level, and at least half of the key muscles below the neurological level have a muscle grade greater than or equal to 3. 
    \item Grade E: Motor and sensory functions are normal 
    \item I do not know/Decline to answer 
\end{itemize}

\noindent\includegraphics[height=4.5in]{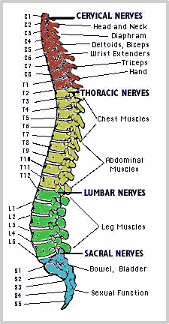}

\noindent 4. If known, what is your level of impairment?
\begin{itemize}[label=$\square$]
    \item C1
    \item C2
    \item C3
    \item C4
    \item C5
    \item C6
    \item C7
    \item C8
    \item T1
    \item T2
    \item T3
    \item T4
    \item T5
    \item T6
    \item T7
    \item T8
    \item T9
    \item T10
    \item T11
    \item T12
    \item L1
    \item L2
    \item L3
    \item L4
    \item L5
    \item S1
    \item S2
    \item S3
    \item S4-5
    \item I do not know/Decline to Answer
\end{itemize}

\vspace{0.25cm}
\hrule
\vspace{0.25cm}

\noindent 5. Please indicate your current level of disability on the scale below. \textit{Please select the description that best describes you.} 
\begin{itemize}[label=$\square$]
    \item 0 - No symptoms 
    \item 1 - No significant disability. Able to carry out all usual activities, despite some symptoms. 
    \item 2 - Slight disability. Able to look after own affairs without assistance, but unable to carry out all previous activities. 
    \item 3 - Moderate disability. Requires some help, but able to walk unassisted. 
    \item 4 - Moderately severe disability. Unable to attend to own bodily needs without assistance, and unable to walk unassisted. 
    \item 5 - Severe disability. Requires constant nursing care and attention, bedridden, incontinent. 
    \item I do not know/Decline to Answer 
\end{itemize}

\vspace{0.25cm}
\hrule
\vspace{0.25cm}

\noindent 6. Please indicate your current type of walking aid. 
\begin{itemize}[label=$\square$]
    \item I use a walking aid such as a cane or leg brace 
    \item I use a wheelchair 
    \item Other 
\end{itemize}

\vspace{0.25cm}
\hrule
\vspace{0.25cm}

\noindent The next few questions are about your opinions on brain-computer interfaces.
\newline

\noindent 7. Have you heard of brain-computer interfaces prior to this survey?
\begin{itemize}[label=$\square$]
    \item Yes 
    \item No 
\end{itemize}

\vspace{0.25cm}
\hrule
\vspace{0.25cm}

\noindent Brain-computer interfaces seek to restore movement or sensation by connecting a person's brain to an external device. This allows a previous motor function or sensation to be restored by bypassing the site of injury and recovering thought control. For example, brain computer interfaces can allow a person to move a robotic arm or regain the sensation of bladder fullness. In the future, we hope to develop a system that is fully implantable.

\noindent\includegraphics[height=3.82in]{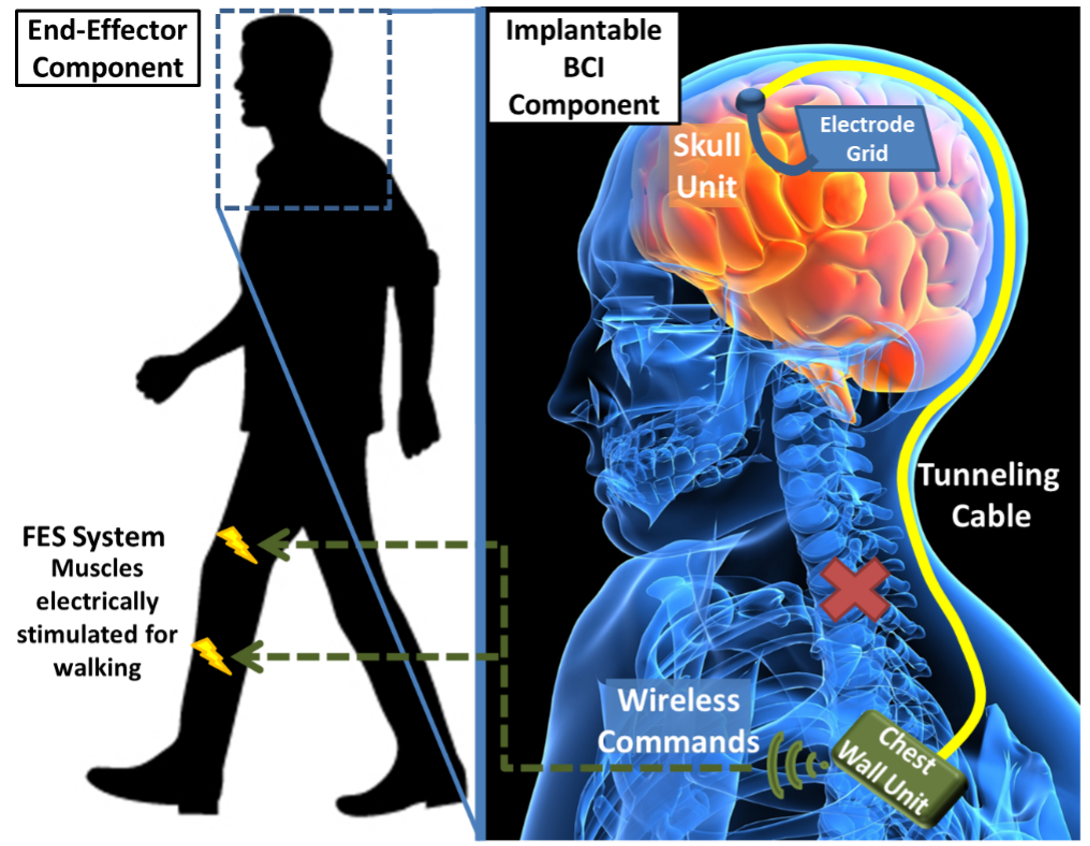}

\noindent 8. How likely are you to consider surgery for a fully implantable brain-computer interface system that would be approximately the size of a pacemaker?
\begin{itemize}[label=$\square$]
    \item Very likely 
    \item Moderately likely 
    \item Slightly likely 
    \item Not at all likely 
\end{itemize}

\vspace{0.25cm}
\hrule
\vspace{0.25cm}

\noindent In order for brain-computer interfaces to work, brain waves must be constantly recorded. One of the current methods to do this is electrocorticography (ECoG). In ECoG, an electrode grid is placed directly on the surface of the brain to record activity from which a person's intentions can be deduced. For this procedure, surgery is required to permanently implant the grid under the skull.

\noindent\includegraphics[height=1.75in]{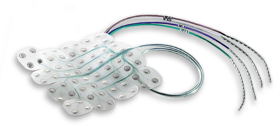} 

\noindent\includegraphics[height=2.11in]{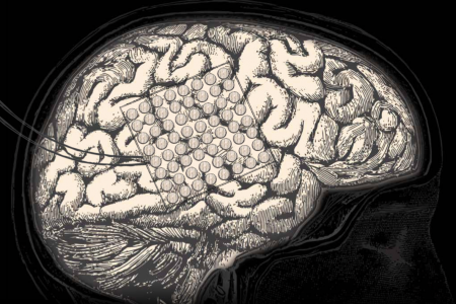} 

\noindent In the next few questions, we are interested in learning whether or not you would be willing to undergo procedures for ECoG implantation if certain motor and sensory functions can be restored.

\vspace{0.25cm}
\hrule
\vspace{0.25cm}

\noindent 9. Please indicate how important regaining arm or upper body function is to you. 
\begin{itemize}[label=$\square$]
    \item Very Important 
    \item Moderately Important 
    \item Slightly Important 
    \item Not At All Important 
    \item Not Applicable 
\end{itemize}

\vspace{0.25cm}
\hrule
\vspace{0.25cm}

\noindent\includegraphics[height=3.5in]{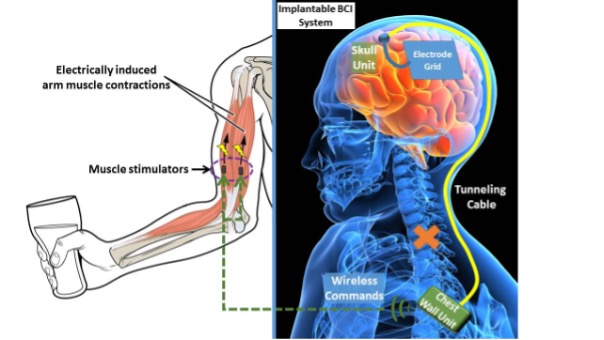} 

\noindent In this setup, an ECoG electrode grid is implanted surgically in the brain to record electrical activity. The signal then gets sent to a chest wall unit where the person's intentions are decoded and translated into commands that are sent wirelessly to implanted muscle stimulators in the arm. The muscles in the arm can therefore be induced to contract and make movements that this person wanted to make. 
\newline

\noindent A similar setup can also be done with a robotic arm, as shown.
\newline

\noindent\includegraphics[height=3in]{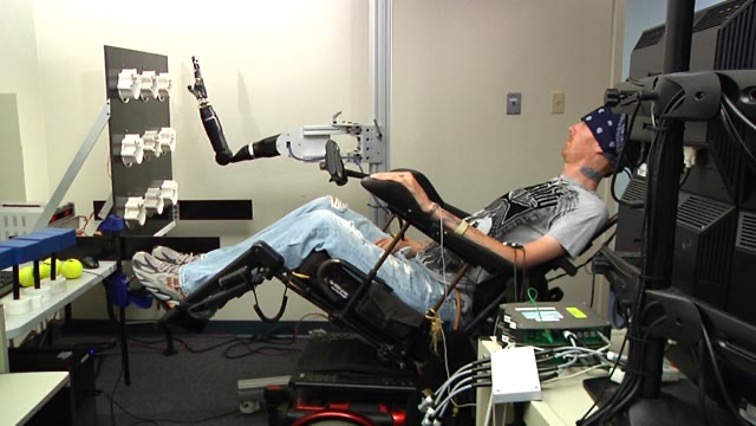}

\noindent 10. How likely would you consider surgery to implant an electrode grid if basic grasp/release ability can be restored?  
\begin{itemize}[label=$\square$]
    \item Very likely 
    \item Moderately likely 
    \item Slightly likely 
    \item Not at all likely 
    \item Not Applicable/Decline to Answer 
\end{itemize}

\noindent 11. How likely would you consider surgery to implant an electrode grid if fine control of fingers, in addition to basic grasp/release ability, can be restored?  
\begin{itemize}[label=$\square$]
    \item Very likely 
    \item Moderately likely 
    \item Slightly likely 
    \item Not at all likely 
    \item Not Applicable/Decline to Answer 
\end{itemize}

\noindent 12. How likely would you consider surgery to implant an electrode grid if fine control of your arm, fine control of your fingers, and basic grasp/release ability can all be restored?  
\begin{itemize}[label=$\square$]
    \item Very likely 
    \item Moderately likely 
    \item Slightly likely 
    \item Not at all likely 
    \item Not Applicable/Decline to Answer 
\end{itemize}

\vspace{0.25cm}
\hrule
\vspace{0.25cm}

\noindent 13. Please indicate how important regaining the ability to walk is to you. 
\begin{itemize}[label=$\square$]
    \item Very Important 
    \item Moderately Important 
    \item Slightly Important 
    \item Not At All Important 
    \item Not Applicable 
\end{itemize}

\vspace{0.25cm}
\hrule
\vspace{0.25cm}

\noindent\includegraphics[height=3.4in]{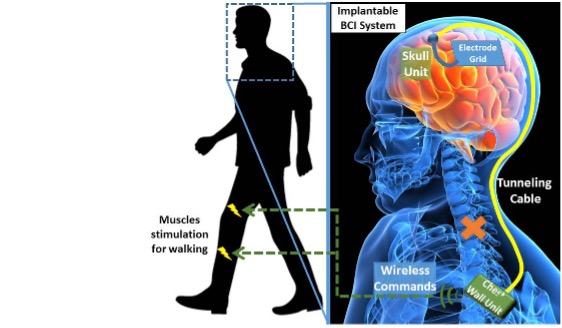}
\newline

\noindent In this setup, an ECoG electrode grid is implanted surgically in the brain to record electrical activity. The signal then gets sent to a chest wall unit where the person's intentions are decoded and translated into commands that are sent wirelessly to implanted muscle stimulators in the leg. The leg muscles can therefore be controlled to make stepping movements for walking.
\newline

\noindent 14. How likely would you consider surgery to implant an electrode grid if you can regain the ability to stand?  
\begin{itemize}[label=$\square$]
    \item Very likely 
    \item Moderately likely 
    \item Slightly likely 
    \item Not at all likely 
    \item Not Applicable/Decline to Answer 
\end{itemize}

\noindent 15. How likely would you consider surgery to implant an electrode grid if you can regain the ability to walk at a constant speed?  
\begin{itemize}[label=$\square$]
    \item Very likely 
    \item Moderately likely 
    \item Slightly likely 
    \item Not at all likely 
    \item Not Applicable/Decline to Answer 
\end{itemize}

\noindent 16. How likely would you consider surgery to implant an electrode grid if you can control walking at various speeds?
\begin{itemize}[label=$\square$]
    \item Very likely 
    \item Moderately likely 
    \item Slightly likely 
    \item Not at all likely 
    \item Not Applicable/Decline to Answer 
\end{itemize}

\noindent 17. How likely would you consider surgery to implant an electrode grid if you can make turns in addition to being able to control walking at various speeds?
\begin{itemize}[label=$\square$]
    \item Very likely 
    \item Moderately likely 
    \item Slightly likely 
    \item Not at all likely 
    \item Not Applicable/Decline to Answer 
\end{itemize}

\vspace{0.25cm}
\hrule
\vspace{0.25cm}

\noindent 18. Please indicate how important regaining sensation is to you. 
\begin{itemize}[label=$\square$]
    \item Very Important 
    \item Moderately Important 
    \item Slightly Important 
    \item Not At All Important 
    \item Not Applicable/Decline to Answer 
\end{itemize}

\vspace{0.25cm}
\hrule
\vspace{0.25cm}

\noindent\includegraphics[height=4.29in]{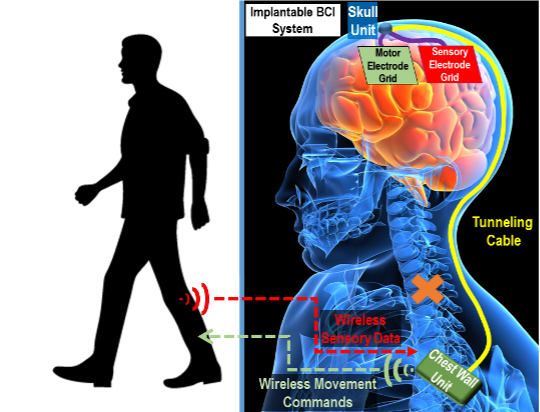}

\noindent In this set-up, an ECoG electrode grid is implanted surgically in the brain to record electrical activity. The signal then gets sent to a chest wall unit where the person's intentions are decoded and translated into commands that are sent wirelessly to implanted muscle stimulators in the leg. The leg muscles can therefore be controlled to make stepping motions for walking. 
\newline

\noindent In addition, sensors placed in the leg can detect when legs are moving and when the feet are placed on the ground. This information is sent to the chest wall unit and converted to signals for the sensory areas of the brain. These signals can provide sensation of leg movements to a person who has lost leg sensation. 
\newline

\noindent Similar set-ups can be used to regain sensation in other parts of the body.

\noindent 19. How likely would you consider surgery to implant an electrode grid if sensation in your legs can be restored?  
\begin{itemize}[label=$\square$]
    \item Very likely 
    \item Moderately likely 
    \item Slightly likely 
    \item Not at all likely 
    \item Not Applicable/Decline to Answer 
\end{itemize}

\noindent 20. How likely would you consider surgery to implant an electrode grid if sensation in your arms can be restored?  
\begin{itemize}[label=$\square$]
    \item Very likely 
    \item Moderately likely 
    \item Slightly likely 
    \item Not at all likely 
    \item Not Applicable/Decline to Answer 
\end{itemize}

\noindent 21. How likely would you consider surgery to implant an electrode grid if sensation in your hands and fingers can be restored?  
\begin{itemize}[label=$\square$]
    \item Very likely 
    \item Moderately likely 
    \item Slightly likely 
    \item Not at all likely 
    \item Not Applicable/Decline to Answer 
\end{itemize}

\vspace{0.25cm}
\hrule
\vspace{0.25cm}

\noindent\includegraphics[width=5in]{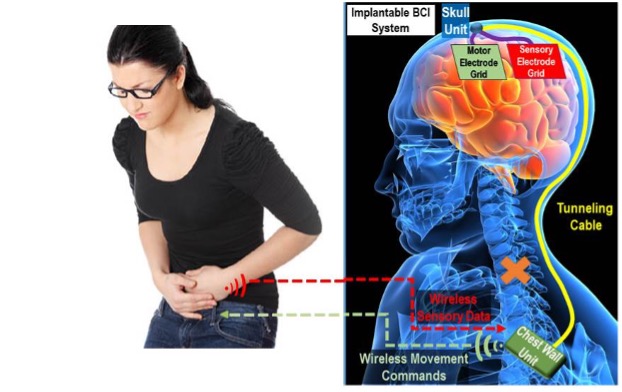}

\noindent In this set-up, an ECoG electrode grid is implanted surgically in the brain to record electrical activity. The signal then gets sent to a chest wall unit where the person's intentions are decoded and translated into commands that are sent wirelessly to implanted muscle stimulators in the sphincters that control urination and defecation. The person can therefore control when they would like to use the restroom. 
\newline

\noindent In addition, sensors placed on the bladder can detect when the bladder is full. This information is sent to the chest wall unit and converted to signals for the sensory areas of the brain. These signals can provide the sensation of bladder fullness to a person who has lost this ability.
\newline

\noindent 22. How likely would you consider surgery to implant an electrode grid if you can regain the sensation of bladder fullness?  
\begin{itemize}[label=$\square$]
    \item Very likely 
    \item Moderately likely 
    \item Slightly likely 
    \item Not at all likely 
    \item Not Applicable/Decline to Answer 
\end{itemize}

\noindent 23. How likely would you consider surgery to implant an electrode grid if you can control when you use the bathroom?  
\begin{itemize}[label=$\square$]
    \item Very likely 
    \item Moderately likely 
    \item Slightly likely 
    \item Not at all likely 
    \item Not Applicable/Decline to Answer 
\end{itemize}

\vspace{0.25cm}
\hrule
\vspace{0.25cm}

\noindent 24. Is there any other bodily function you would like to restore that has not been mentioned here?
\begin{itemize}[label=$\square$]
    \item Yes
    \item No
\end{itemize}

\vspace{0.25cm}
\hrule
\vspace{0.25cm}

\noindent 25. Please explain what motor function(s) or sensation you would like to restore. 
\newline

\noindent \framebox[6 cm][l]{Text box}
\newline

\vspace{0.25cm}
\hrule
\vspace{0.25cm}

\noindent 26. How likely would you consider surgery to implant electrodes in the brain if the function you indicated can be restored?
\begin{itemize}[label=$\square$]
    \item Very likely 
    \item Moderately likely 
    \item Slightly likely 
    \item Not at all likely 
\end{itemize}

\vspace{0.25cm}
\hrule
\vspace{0.25cm}

\noindent 27. Can you please explain why you would not be interested in surgery for electrode implantation in the brain if the desired function can be restored? 
\newline

\noindent \framebox[6 cm][l]{Text box}
\newline

\vspace{0.25cm}
\hrule
\vspace{0.25cm}

\noindent 28. What kind of concerns would you have regarding a surgery to implant a BCI system? Select all answers that apply.
\begin{itemize}[label=$\square$]
    \item Cost of surgery for BCI implantation
    \item Potential risks and possible complications of surgery such as infection, scarring, excessive bleeding, blood clots, reactions to anesthetics
    \item Motor and/or sensory restoration not meeting expectations for practical use
    \item Length and effort of training required to fully maximize BCI usage
    \item Possible need for additional surgical interventions if BCI fails
    \item Long-term usage and durability of device
    \item None
    \item Other
\end{itemize}

\vspace{0.25cm}
\hrule
\vspace{0.25cm}

\noindent 29. (Optional) Please indicate any other concern you have about surgery for electrode implantation or BCI technology that is not listed above.
\newline

\noindent \framebox[6 cm][l]{Text box}
\newline

\vspace{0.25cm}
\hrule
\vspace{0.25cm}

\noindent Finally, we would like to know more about you, to help compare your answers to those of other participants.  Again, all of this information is confidential.
\newline

\noindent 30. What is your age?
\begin{itemize}[label=$\square$]
    \item 18-24 years old 
    \item 25-34 years old 
    \item 35-44 years old 
    \item 45-54 years old 
    \item 55-64 years old 
    \item 65-74 years old 
    \item 75 years or older 
    \item I do not know/Decline to Answer
\end{itemize}

\noindent 31. I am:
\begin{itemize}[label=$\square$]
    \item Male 
    \item Female 
    \item Other 
    \item Decline to Answer 
\end{itemize}

\noindent 32. Please indicate the highest level of education you have completed:
\begin{itemize}[label=$\square$]
    \item Some high school or less 
    \item High School Diploma or equivalent 
    \item Some college 
    \item College graduate 
    \item Advanced Degree (Master's, JD, MD, etc.) 
    \item I do not know/Decline to Answer
\end{itemize}

\noindent 33. Please indicate your current occupation setting:
\begin{itemize}[label=$\square$]
    \item Laborer or helper (examples: grounds maintenance worker, construction laborer) 
    \item Operative (examples: machine operator, parking lot attendant, bus driver) 
    \item Craft worker (examples: electrician, plumber, construction worker, painter) 
    \item Service worker (examples: cook, food preparation worker, custodian) 
    \item Security support (examples: police officer, security guard) 
    \item Commercial/sales support (examples: sales supervisor, cashier, travel agent) 
    \item Medical support (examples: medical assistant, healthcare worker) 
    \item Administrative support (examples: office manager, library technician, secretary, payroll clerk, accounting assistant) 
    \item Technician (examples: laboratory technician, LPN, diagnostic related technologist) 
    \item Professional (examples: instructor, engineer, scientist, physician, pharmacist, registered nurse, librarian, computer programmer, HR specialist, accountant, financial analyst, athletic coach) 
    \item Manager or official (executive officer, mid-level manager) 
    \item Other 
    \item Don't know
\end{itemize}

\noindent 34. What is your current living situation?
\begin{itemize}[label=$\square$]
    \item I live in a skilled nursing facility with health care staff providing full-time assistance
    \item I live in an assisted living facility with health care staff providing partial assistance
    \item I live at home, but use supportive living services (a health care worker provides assistance when needed)
    \item I live at home, but somebody else (family, friend, caregiver) helps me complete daily tasks
    \item I live at home and am able to complete daily tasks without assistance
    \item Don't know/Decline to answer
\end{itemize}

\noindent 35. What is your household’s annual income?  Include earnings, Social Security, disability payments, and any other type of income. 
\begin{itemize}[label=$\square$]
    \item \$9,999 or less 
    \item \$10,000-\$19,999 
    \item \$20,000-\$39,999 
    \item \$40,000-\$59,999 
    \item \$60,000-\$99,999 
    \item \$100,000-\$199,999 
    \item \$200,000 or higher 
    \item I do not know/Decline to Answer 
\end{itemize}

\vspace{0.25cm}
\hrule
\vspace{0.25cm}

\vspace{0.25cm}
\hrule
\vspace{0.25cm}

\begin{center}
You have reached the end. Thank you for participating.
\end{center}

\vspace{0.25cm}
\hrule

\begin{center}
Bottom of Form
\end{center}

\newpage

\section{Additional Figures}
\setcounter{figure}{0} 
\renewcommand{\thefigure}{a\arabic{figure}}

\begin{figure*}[!htbp]
    \centering  
    \includegraphics[width=5in]{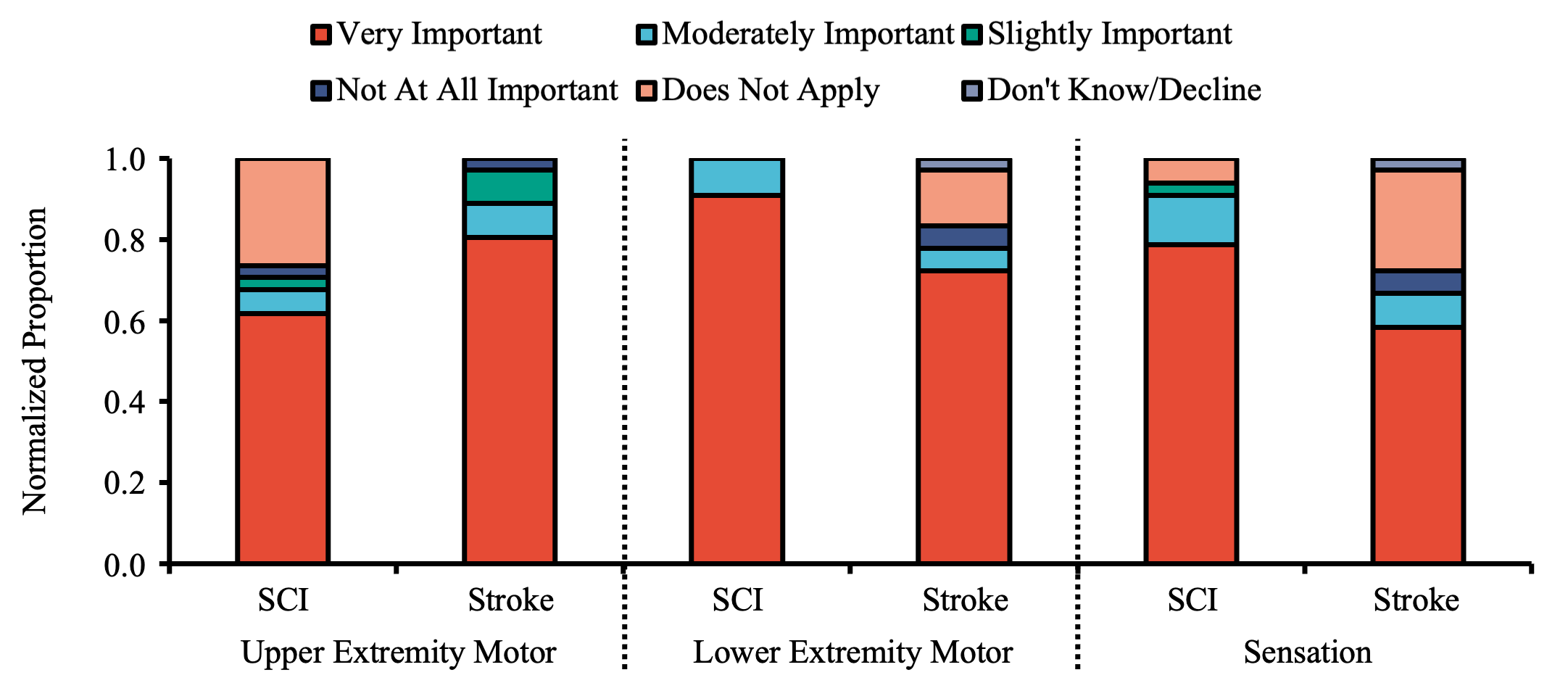} 
    \caption{Importance of regaining motor and sensory functions to stroke and SCI participants as a normalized proportion. The majority ($> 60\%$) of both stroke and SCI participants indicated that regaining upper extremity motor function, lower extremity motor function, and extremities and bladder sensation were all ``Very Important." }
    \label{fig:participant_perception_sup}
\end{figure*}

\begin{figure*}[!htbp]
    \centering
    \includegraphics[width=5in]{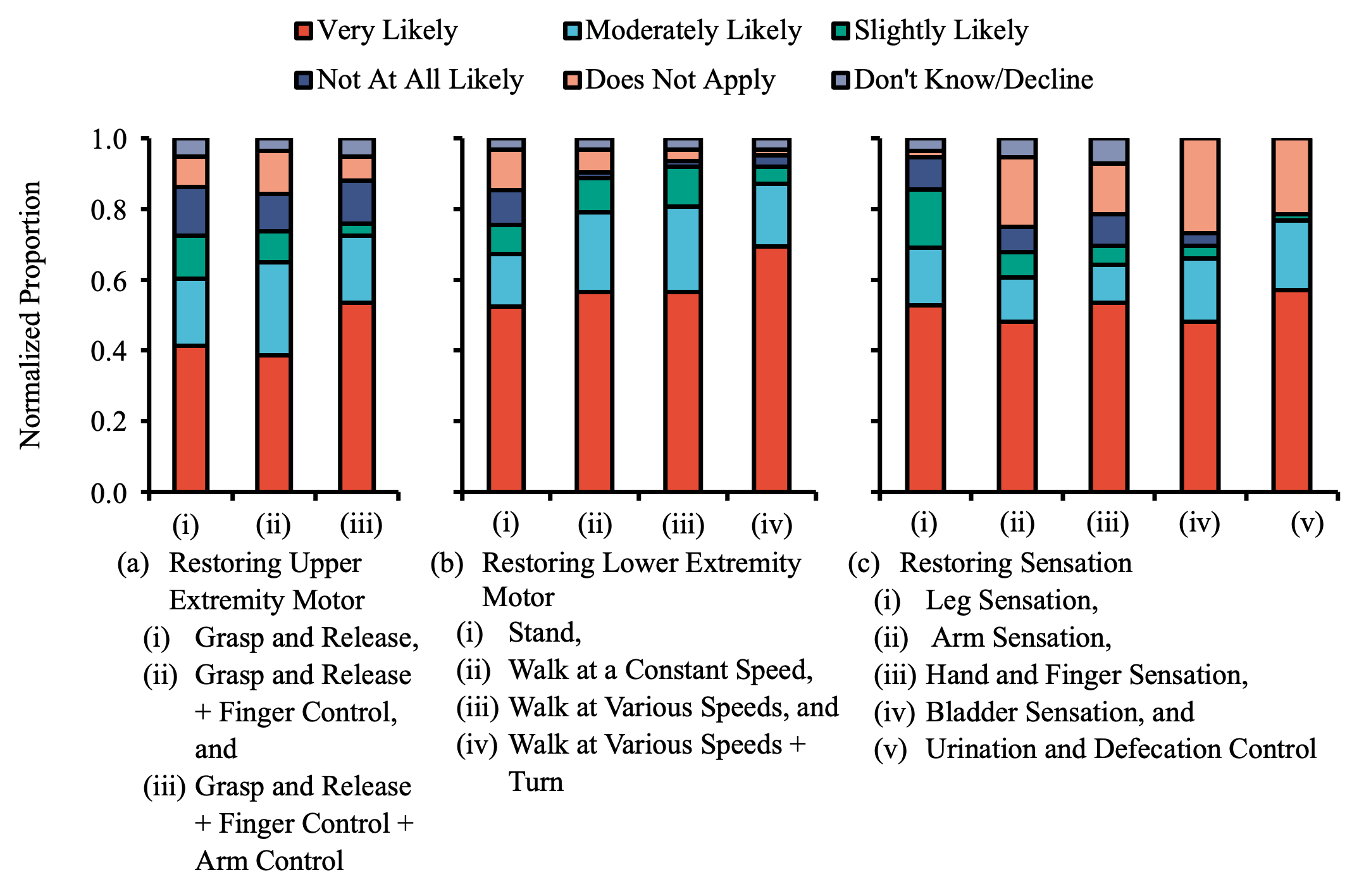}
    \caption{Participant willingness to undergo surgery to implant BCI system at various degrees of function restored. (a) Upper extremity motor function. (b) Lower extremity motor function. (c) Sensation. Participants were asked to indicate their level of disability as well as how important regaining upper extremity motor function, lower extremity motor function, and sensation were to them. Participants’ responses were plotted as a sample size-normalized stacked bar graph of those who indicated that each function was ``very important,'' ``moderately important,'' ``slightly important,'' ``not at all important,'' or ``does not apply.'' The majority of participants were at least moderately willing to undergo surgery to implant BCI systems for even basic levels of functional restoration.}
    \label{fig:willingness_to_undergo_surgery_sup}
\end{figure*}

\begin{figure*}[!htbp]
    \centering  
    \includegraphics[width=5in]{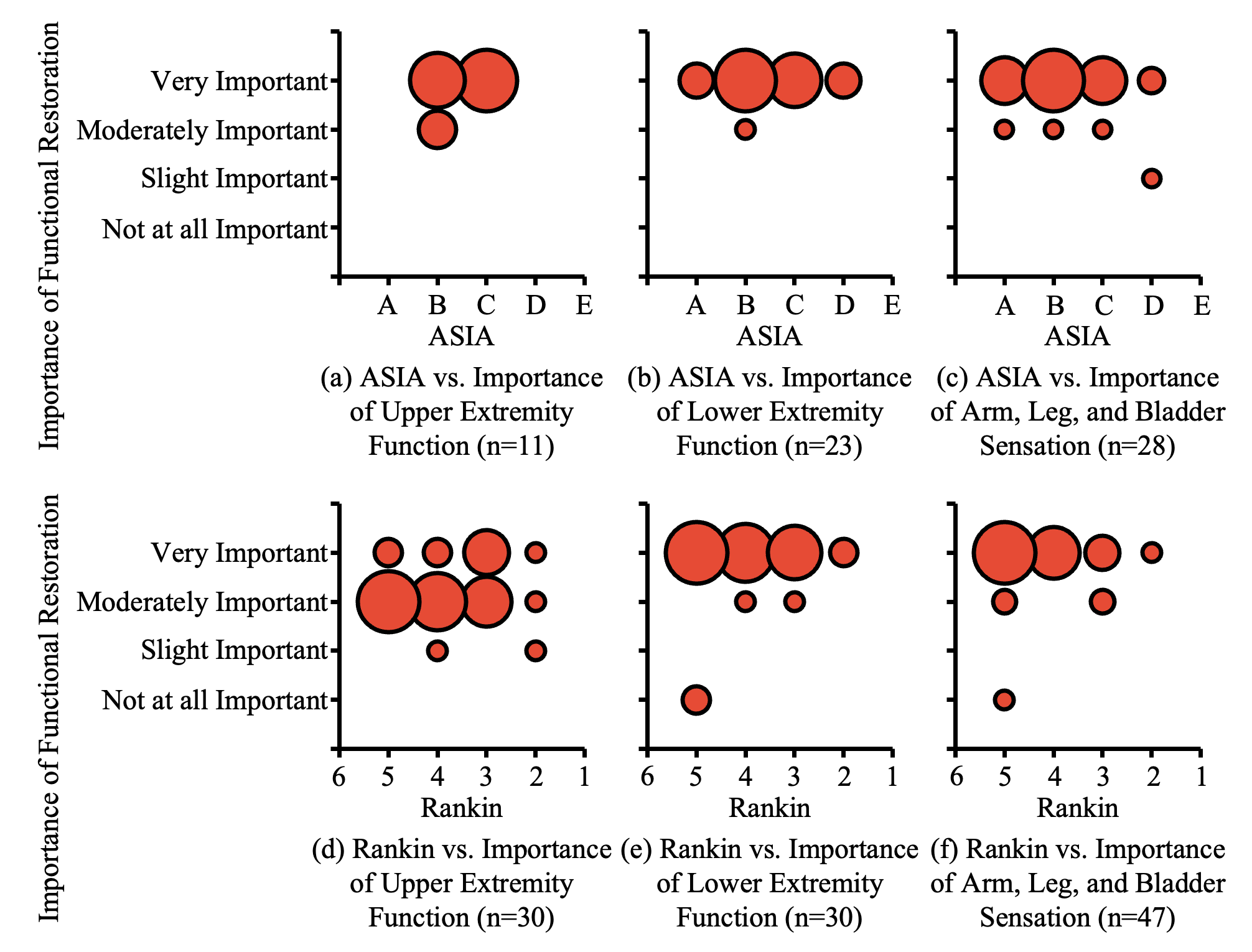}
    \caption{Importance of functional restoration given different levels of disability measured by ASIA Impairment or Modified Rankin Scale. ASIA scale ranks disability from A to E, with A being no motor, sensory or sacral sparing and E being normal motor and sensory function. Modified Rankin goes from 0 to 6 with 0 being no disability and 6 being dead. ``Does not apply'' responses were excluded.}
    \label{fig:importance_of_functional_restoration}
\end{figure*}

\begin{figure*}[htbp]
    \centering  
    \includegraphics[width=5in]{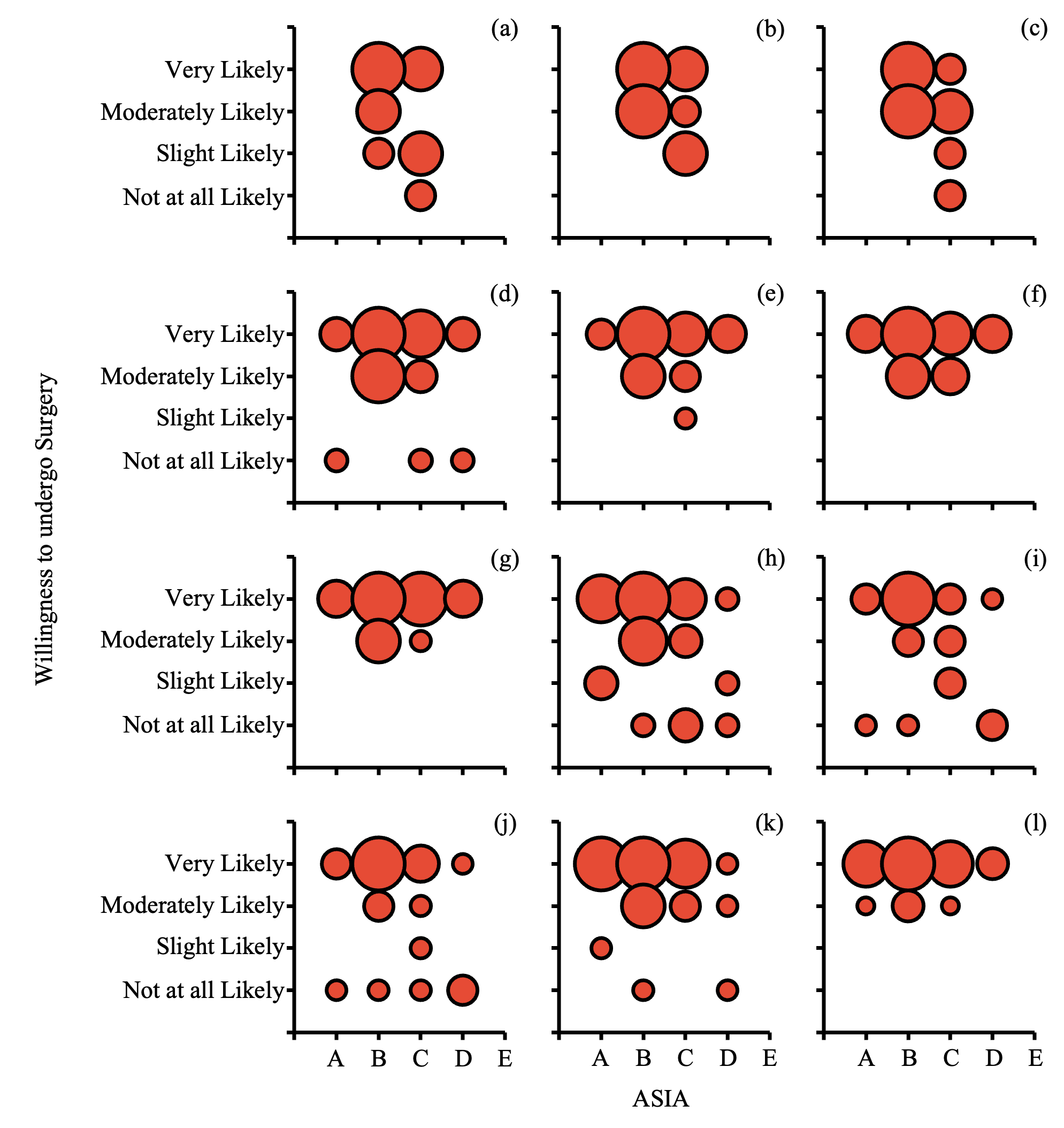}
    \caption{
        Participants' willingness to undergo surgery to restore different functions given different levels of disability measured by AISA Impairment Scale. ASIA scale ranks disability from A to E, with A being no motor, sensory or sacral sparing and E being normal motor and sensory function. “does not apply” was excluded from the bubble charts. 
        ASIA vs. Restoring 
        (a) Grasp and Release (n=11);
        (b) Grasp and Release + Finger Control (n=11);
        (c) Grasp and Release + Finger Control + Arm Control (n=11).
        ASIA vs. Restoring Ability to 
        (d) Stand (n=23);
        (e) Walk at Constant Speed (n=22);
        (f) Walk at Various Speeds (n=23);
        (g) Walk at Various Speeds (f) + Turn (n=23);
        ASIA vs. Restoring 
        (h) Leg Sensation (n=26);
        (i) Arm Sensation (n=21);
        (j) Hand and Finger Sensation (n=21);
        (k) Bladder Fullness Sensation (n=28);
        (l) Urination and Defecation Control (n=28);
    }
    \label{fig:asia}
\end{figure*}

\begin{figure*}[!htbp]
    \centering
    \includegraphics[width=5in]{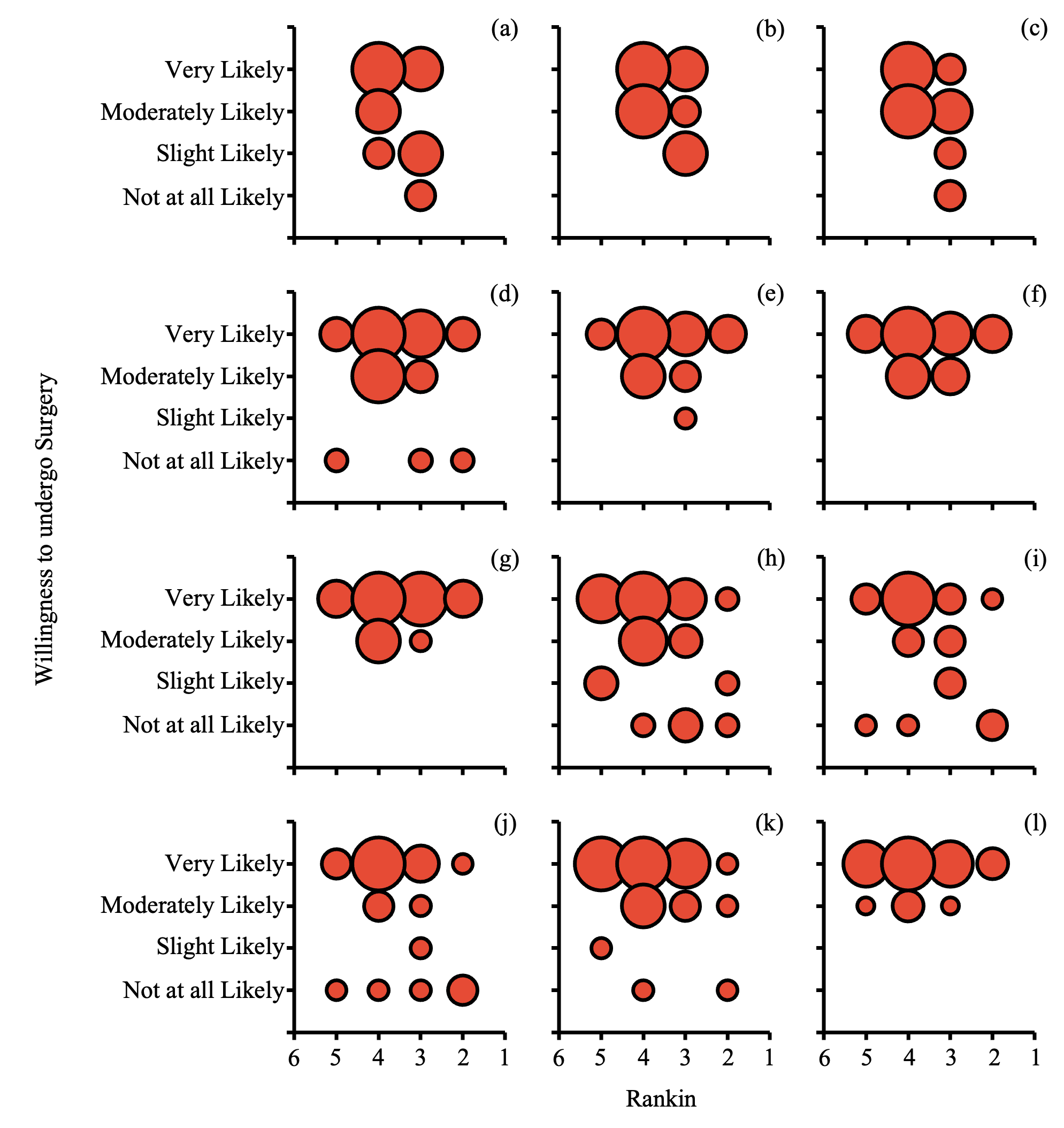}
    \caption{
        Participant's willingness to undergo surgery to restore different functions given different levels of disability measured by Modified Rankin Scale. Modified Rankin goes from 0 to 6 with 0 being no disability and 6 being dead. “does not apply” was excluded from the bubble charts.
        Rankin vs. Restoring 
        (a) Grasp and Release (n=28);
        (b)  Grasp and Release + Finger Control (n=28);
        (c) Grasp and Release + Finger Control + Arm Control (n=30).
        Rankin vs. Restoring Ability to 
        (d) Stand (n=21);
        (e) Walk at Constant Speed (n=24);
        (f) Walk at Various Speeds (n=24);
        (g) Walk at Various Speeds + Turn (n=26).
        Rankin vs. Restoring 
        (h) Leg Sensation (n=43);
        (i) Arm Sensation (n=33);
        (j) Hand and Finger Sensation (n=35);
        (k) Bladder Fullness Sensation (n=31);
        (l) Urination and Defecation Control (n=34).
    }
    \label{fig:rankin}
\end{figure*}

\begin{figure*}[!htbp]
    \centering
    \includegraphics[width=5in]{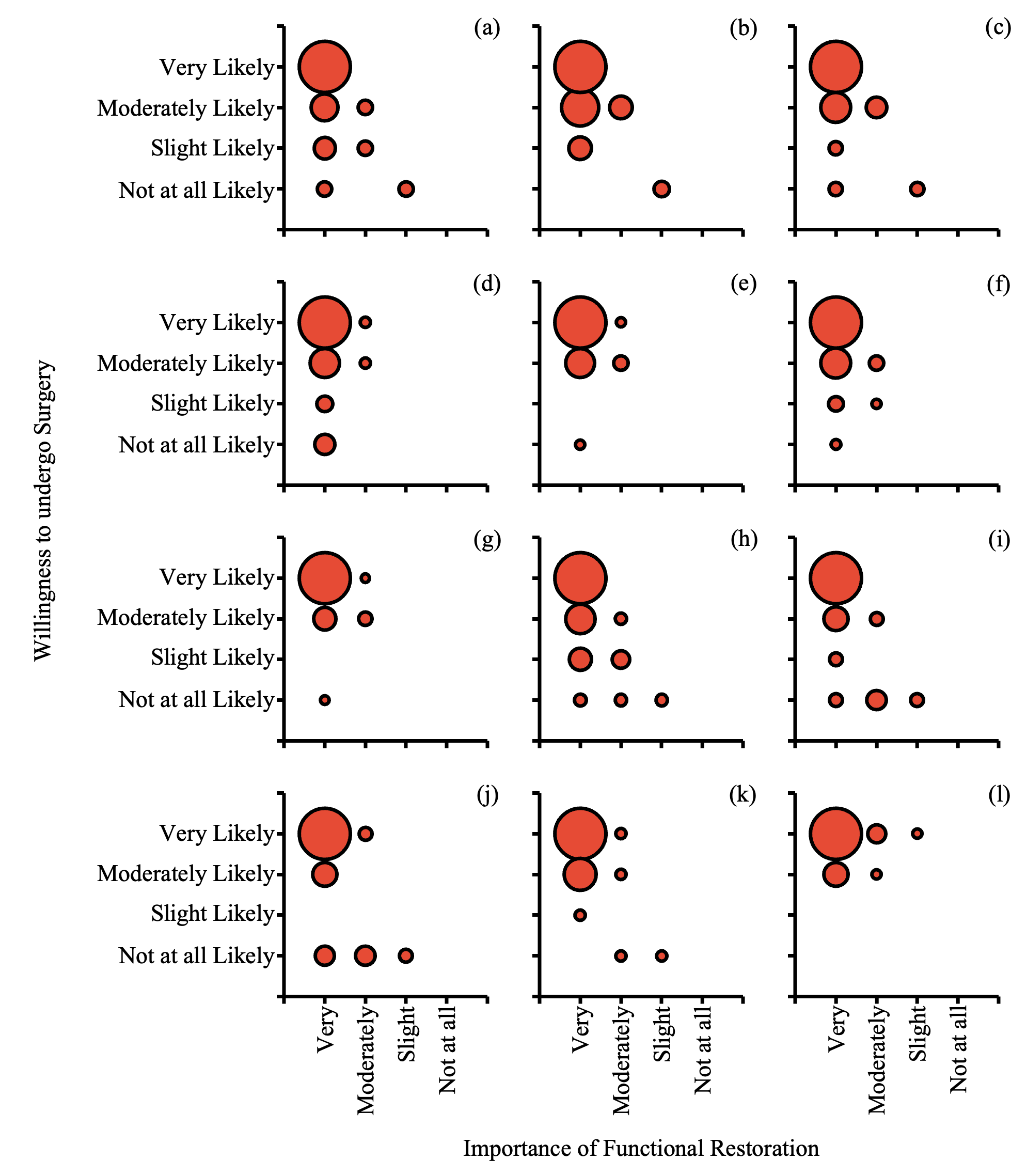}
    \caption{
        SCI Participant's willingness to undergo surgery to restore different functions given Perceived Importance of Functional Restoration. “does not apply” was excluded from the bubble charts. 
        Importance of Upper Extremity Function vs. Restoring 
        (a) Grasp and Release (n=19); 
        (b) Grasp and Release + Finger Control (n=19);
        (c) Grasp and Release + Finger Control + Arm Control (n=20).
        Importance of Lower Extremity Function vs. Restoring Ability to 
        (d) Stand (n=30);
        (e) Walk at Constant Speed (n=31);
        (f) Walk at Various Speeds (n=32);
        (g) Walk at Various Speeds + Turn (n=32).
        Importance of Sensation vs. Restoring
        (h) Leg Sensation (n=28);
        (i) Arm Sensation (n=21);
        (j) Hand and Finger Sensation (n=21);
        (k) Bladder Fullness Sensation (n=29);
        (l) Urination and Defecation Control (n=30).
    }
    \label{fig:sci_restoration_vs_sx}
\end{figure*}

\begin{figure*}[!htbp]
    \centering
    \includegraphics[width=5in]{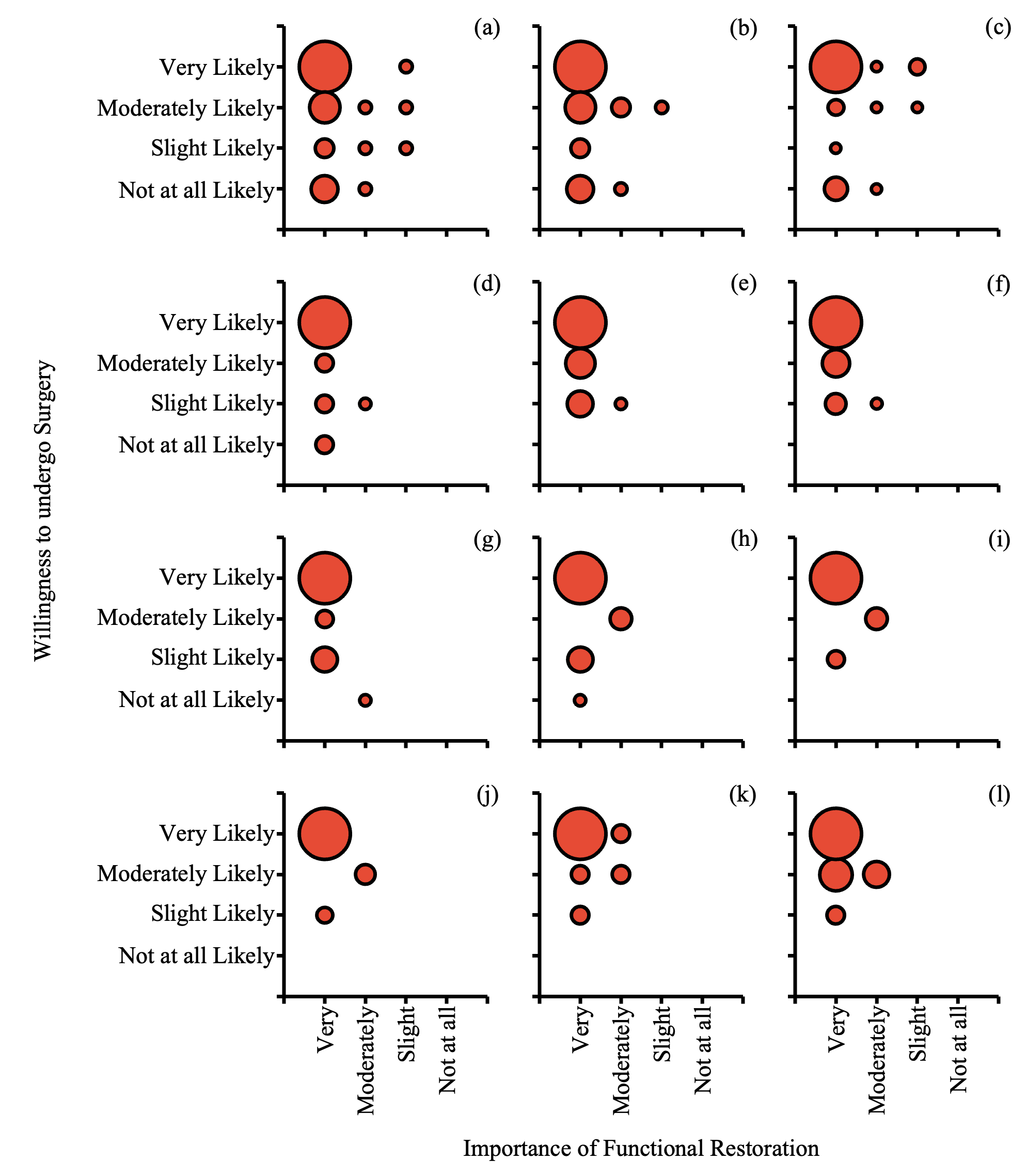}
    \caption{
        Stroke participant's willingness to undergo surgery to restore different functions given Perceived Importance of Functional Restoration. “does not apply” was excluded from the bubble charts.
        Importance of Upper Extremity Function vs. Restoring 
        (a) Grasp and Release (n=30);
        (b) Grasp and Release + Finger Control (n=28);
        (c) Grasp and Release + Finger Control + Arm Control (n=30).
        Importance of Lower Extremity Function vs. Restoring Ability to 
        (d) Stand (n=21);
        (e) Walk at Constant Speed (n=24);
        (f) Walk at Various Speeds (n=25);
        (g) Walk at Various Speeds + Turn (n=22).
        Importance of Sensation vs. 
        (h) Leg Sensation (n=23);
        (i) Arm Sensation (n=20);
        (j) Hand and Finger Sensation (n=22);
        (k) Bladder Fullness Sensation (n=11);
        (l) Urination and Defecation Control (n=13).
    }
    \label{fig:stroke_restoration_vs_sx}
\end{figure*}

\end{appendices}

\end{document}